\documentclass[a4paper]{article}

\usepackage[left=1.5cm, right=1.5cm, top=1.5cm, bottom=1.5cm]{geometry}

\usepackage{amsmath,amssymb}

\usepackage{url}

\usepackage[british]{babel}
\usepackage[utf8]{inputenc}
\usepackage{graphicx}
\usepackage{xcolor}
\usepackage{xspace}
\usepackage{xfrac}
\usepackage{tikz}
\usepackage{pgfplotstable}
\pgfplotsset{compat=1.13}
\usepgfplotslibrary{fillbetween}

\usepackage{lineno}

\newtheorem{definition}{Definition}

\newcommand{\rd}{\mathrm{d}}

\newcommand{\e}{\mathrm{e}}

\newcommand{\R}{\mathbb{R}}
\newcommand{\bF}{\mathbf{F}}
\newcommand{\bM}{\mathbf{M}}

\newcommand{\bx}{\mathbf{x}}

\newcommand{\ess}{\mathrm{ESS}}
\newcommand{\eff}{\mathrm{eff}}
\newcommand{\X}{\ensuremath{\mathbb{X}}\xspace}
\newcommand{\A}{\ensuremath{\mathbb{A}}\xspace}
\newcommand{\B}{\ensuremath{\mathbb{B}}\xspace}
\newcommand{\C}{\ensuremath{\mathbb{C}}\xspace}
\newcommand{\Z}{\ensuremath{\mathbb{Z}}\xspace}
\newcommand{\Vmean}{\ensuremath{\langle V\rangle_\kappa}}
\newcommand{\Neff}{\ensuremath{N_{\mathrm{eff}}}}

\newcommand{\bydef}{:=}
\newcommand{\defby}{=:}

\graphicspath{{figs/}}

\begin{document}

%\linenumbers

\title{Fitness potentials and qualitative properties of the Wright-Fisher dynamics}

\author{Fabio A. C. C. Chalub\textsuperscript{1} \and Max O. Souza\textsuperscript{2,3}}

\date{}

\maketitle

\footnotetext[1]{Departamento de Matemática and Centro de Matemática e Aplicações, Faculdade de Ciências e Tecnologia, Universidade Nova de Lisboa, Quinta da Torre, 2829-516, Caparica, Portugal.}

\footnotetext[2]{Instituto de Matemática e Estatística, Universidade Federal Fluminense, R. Prof. Marcos Waldemar de Freitas Reis, s/n, 24210-201, Niterói, RJ, Brasil.}

\footnotetext[3]{E-mail: maxsouza@id.uff.br}

\begin{abstract}
	We present a mechanistic formalism for the study of evolutionary dynamics models based on the diffusion approximation described by the Kimura Equation. In this formalism, the central component is the fitness potential, from which we obtain an expression for the amount of work necessary for a given type to reach fixation. In particular, within this interpretation, we develop a graphical analysis --- similar to the one used in classical mechanics --- providing the basic tool for a simple heuristic that describes both the short and long term dynamics. As a by-product, we provide a new definition of an evolutionary stable state in finite populations that includes the case of mixed populations. We finish by showing that our theory -- rigorous for two types evolution without mutations-- is also consistent with the multi-type case, and with the inclusion of rare mutations.
	
\end{abstract}

\section{Background}
 
Out of many properties of evolutionary models for finite populations, the fixation probability of a given type as a function of the current state of the population is likely  the most important one, and methods guaranteeing  its correct estimation are important~\cite{Imhof:Nowak:2006}. Among these models, the Moran (M)~\cite{Moran} and Wright Fisher (WF)\cite{Fisher1,Wright2} processes are  archetypal and share many properties in common --- but see~\cite{ChalubSouza17a} for important differences also. In the regime of large population and weak selection, both processes share the same diffusion approximation, namely the Kimura Equation (KE).
From now on, for the sake of simplicity,  we will refer only to the WF process when discussing evolution in finite populations. Nevertheless, all results discussed here will apply equally well to any Markov chain whose diffusion approximation is given by the Kimura Equation~\cite{ChalubSouza09a,ChalubSouza14a}. We will develop a graphical procedure that allows to obtain many qualitative features	 of the fixation probability and, by extension, of the full history of a population evolving according to the WF process. This procedure resembles the use of potentials and energies in classical mechanics~\cite{Goldstein}. Using what we call the \emph{fitness potential} we are able to qualitatively analyse in an unified way the WF process, the Replicator Equation (RE) and the Kimura Equation;  
the last one collapses several parameters of the WF processes in one new parameter, the \emph{effective population size} $\Neff=\kappa^{-1}$~\cite{Charlesworth}.

Convergence of the WF process  towards the KE, when the population size goes to infinity,  has been shown under various assumptions --- cf.~\cite{EthierKurtz,ChalubSouza14a}. Additionally, the estimates in \cite{ChalubSouza09b} suggest that the
KE approximates the WF process over all time scales, in the limit of large population size and weak selection; furthermore, when $\kappa\ll1$, the short time dynamics of the KE is well approximated by solutions of the (PDE version of) RE \cite{ChalubSouza14a}; this approximation is, in general, non-uniform in time. Thus, in this regime, one might envisage a two-stage dynamics for the WF process: an RE phase, and a post RE phase. Our aim is then to describe both phases using two different interpretations of the  fitness potential graph.

For the two types (2T) case, the fitness potential $V$ introduced in the context of the Moran process~\cite{ChalubSouza16a} is a function of the presence of the focal type (also called the population state), such that its derivative equals the fitness difference between the focal and opponent types, i.e. $-V'$ is the gradient of selection. With the diffusion approximation for fixation in mind, we introduce
the (generalized Kolmogorov $\kappa/2$-)exponential mean of $V$, $\Vmean$~\cite{Carvalho_2016}. Intuitively, the population state faces difficulties to reach potential (energy) levels of $V$ above $\Vmean$ and therefore in the two types case the population will eventually fixate with higher probability in the vertex that can be reached with minimum energy. In the multi-type case, the first type to be extinct will be the one opposed to the face that can be reached with minimum energy. When no face can be reached with energy levels below $\Vmean$, then the system takes significantly longer times in the interior of the simplex; this motivates the definition of $\kappa$-evolutionary stable state $\ess_\kappa$ (akin to $\ess_N$, see~\cite{Nowak:06}, but focusing on the effective population size and not on the real population size) for non-monomorphic populations. 

It worth pointing out that, while this  analysis holds for fairly arbitrary gradients of selection, special attention will be given to those that arise from Evolutionary Game Theory (EGT), as first introduced in \cite{smith1973logic}.  These are particularly relevant to applications in population dynamics,  as shown by the several contributions of  EGT either to the study of interactions in a population \cite{may1994superinfection,may1995coinfection,kerr2002local} or to cultural evolution \cite{nowak2002computational,trivers1971evolution,axelrod1981evolution,boyd2005origin,nowak2005evolution} among many others.

When mutations are included, but are very rare, the long term dynamics spends a significant fraction on the monomorphic states --- see \cite{fudenberg2006imitation} and references therein. However, this does not mean that its long-term statistics are necessarily the same as for the process with no mutations \cite{wu2012small}. Indeed, numerical simulations suggest that when there is an interior point $x_*$ which is an $\ess_\kappa$ of the finite population dynamics, then the quasi-stationary distribution of the non-mutation case will be similar to the small mutation case; this will not be true if the interior point $x_*$ is a stable equilibrium of the deterministic dynamics, but not an $\ess_\kappa$. 

Numerical results  were obtained using standard Markov chain algorithms \cite{Karlin_Taylor_first}.

\section{Methods}
\label{sec:fixation_energy}

\subsection{Wright-Fisher process}

Consider a population of $N$ haploid individuals consisting of types \A and \B. There are no mutations.
The WF process is a Markov chain with state space $\lbrace i/N\,:\, i=0,\ldots,N\rbrace$, where $i$ denotes the presence of type \A individuals in the population, and with transition probabilities given by the row stochastic matrix $\bM=\left(M_{ij}\right)_{i,j=0\dots,N}$,  $M_{ij}=\binom{N}{j}p_i^j(1-p_i)^{N-j}$,  $p_0=1-p_N=0$ and $p_i\in(0,1)$ for $i=1,\dots,N-1$. The fixation probability of type \A when the population is initially at state $i/N$ is denoted $F_i$. The vector $\bF^\dagger=(0,F_1,\dots,F_{N-1},1)$ is the unique solution of $\bF=\bM\bF$, $F_0=1-F_N=0$~\cite{Karlin_Taylor_first}.

Time for fixation can be obtained in a similar way: Let $T_i$ be the time for type \A or type \B be fixed (i.e., the time such that the system reaches a absorbing state), when the population is initially at state $i$. Therefore $T_i=\sum_jM_{ij}T_j+1$, with $T_0=T_N=0$. In particular, following the notation of~\cite{ChalubSouza17a}, $\widetilde{\bM}=\left(M_{ij}\right)_{i,j=1,\dots,N-1}$, $\widetilde{\bF}=(F_1,\dots,F_{N-1})^\dagger$, $\mathbf{b}=(M_{1N},\dots,M_{N-1,N})^\dagger$, $\mathbf{T}=(T_0,T_1,\dots,T_N)^\dagger$, $\widetilde{\mathbf{T}}=(T_1,\dots,T_{N-1})^\dagger$. Finally, $\widetilde{\bF}=\left(I-\widetilde{\bM}\right)^{-1}\widetilde{\mathbf{b}}$ and $\widetilde{\mathbf{T}}=\left(\mathbf{I}-\widetilde{\mathbf{M}}\right)^{-1}\mathbf{1}$, where $\mathbf{I}$ is the $(N-1)\times(N-1)$ unity matrix and $\mathbf{1}=(1,1,\dots,1)^\dagger\in\mathbb{R}^{N-1}$.

For each type, we define fitnesses functions $\Psi^{(\A),(\B)}:\{0,\dots,N\}\to\R_+$, such that the probability of selection for reproduction is given by  
\begin{equation}\label{eq:piweakselection}
 p_i=\frac{i\Psi^{(\A)}(i)}{i\Psi^{(\A)}(i)+(N-i)\Psi^{(\B)}(i)}\ .
\end{equation}

\subsection{Diffusion approximation}

 In the weak selection approximation for large populations~\cite{ChalubSouza14a}, we write $\Psi^{(\X)}(i)=1+\left(\kappa N\right)^{-1}\psi^{(\X)}(i/N)$, with $\psi^{(\X)}:[0,1]\to\R$, and $\kappa N$ sufficiently large such that $\Psi^{(\X)}$ is always positive. Then, we have that
\[
p_i=\frac{i}{N}\left[1-\frac{1}{\kappa N}\frac{N-i}{N}V'\left(\frac{i}{N}\right)\right]+\mathrm{o}\left[\left(\kappa N\right)^{-1}\right]\ .
\]
where $V(x)=-\int_0^x\theta(y)\rd y$ is the fitness potential and $\theta=\psi^{(\A)}-\psi^{(\B)}$ is the fitness difference between types \A and \B.

The fixation probability $F_i$ can then be approximated by 
\begin{equation}\label{eq:fix_prob_continous}
\varphi(x)=\frac{\int_0^x\e^{\frac{2}{\kappa}V(y)}\rd y}{\int_0^1\e^{\frac{2}{\kappa}V(y)}\rd y}=\int_0^x\e^{\frac{2}{\kappa}\left(V(y)-\Vmean\right)}\rd y\ ,
\end{equation}
where $x=i/N$, and 
\begin{equation*}
\Vmean=\frac{\kappa}{2}\log\int_0^1\e^{\frac{2}{\kappa}V(y)}\rd y\ 
\end{equation*}
is the $\dfrac{2}{\kappa}$-exponential average of $V$, i.e., $\Vmean=\varphi^{-1}\left(\int_0^1\varphi(V(y))\rd y\right)$, for $\varphi(x)=\e^{2 x/\kappa}$; see~\cite{Ewens_Book}  for a classical presentation and~\cite{ChalubSouza09a,ChalubSouza16a} for specific results on the Moran process.

\section{Results}

\subsection{Work and probability of fixation}

Let $\rho(x)$ be the the ratio between the fixation probability of types \A and \B as a function of the presence of type \A, i.e., $\rho(x)=\frac{\varphi(x)}{1-\varphi(x)}$. A direct calculation yields
\begin{equation*}
 \rho(x)=\frac{\int_0^x\e^{\frac{2}{\kappa}(V(y)-\Vmean)}\rd y}{\int_x^1\e^{\frac{2}{\kappa}(V(y)-\Vmean)}\rd y}=\left\{\left[\int_0^x\e^{\frac{2}{\kappa}(V(y)-\Vmean)}\rd y\right]^{-1}-1\right\}^{-1}\ .
\end{equation*}

Note that $\int_x^1\e^{\frac{2}{\kappa}(V(y)-\Vmean)}\rd y$ may be interpreted as the work necessary to transport type \A individuals from the initial state until fixation. 

Rewriting the fixation problem from the point of view of the type \B, we have that the fitness difference between types \B and \A is $\theta^{(\B)}(1-x)=-\theta(x)$. Thus, if we denote the presence of type \B by $x'=1-x$, then $\theta^{(\B)}(x')=-\theta(1-x')$. The corresponding fitness potential is given by 
\[
 V^{(\B)}(x')=-\int_{0}^{x'}\theta^{(\B)}(y)\rd y=\int_0^{x'}\theta(1-y)\rd y=\int_{1-x'}^1\theta(y)\rd y=-V(1)+V(1-x')\ .
\]
Therefore $\langle V^{(\B)}\rangle_\kappa=-V(1)+\Vmean$, and the necessary work to fixate type \B is given by 
\[
\int_{x'}^1\e^{\frac{2}{\kappa}(V^{(\B)}(y)-\langle V^{(\B)}\rangle_\kappa)}\rd y=\int_0^x\e^{\frac{2}{\kappa}(V(y)-\Vmean)}\rd y\ .
\]
We conclude that the ratio of the fixation probabilities of both types is the inverse of the ratio of the work necessary for each type to reach fixation. This  will be a key fact for the mechanical analogy that we are going to develop in the remainder of this work.

\subsection{Graphical analysis}
\label{sec:graphical}

When considering the KE with parameter $\kappa\ll1$, we are assuming selection-dominated evolution, i.e., the genetic drift is assumed to be weak. As already stated, the KE is a uniform in time approximation for the WF dynamics in the limit of large populations and weak selection. In this scenario, the earlier  WF dynamics is well approximated by the RE~\cite{ChalubSouza14a}. However, for longer times, diffusion, representing genetic drift, prevails and the population will eventually reach a monomorphic state, i.e.,  $x=0$ or $x=1$.  With this dichotomy in mind,  we will show how a simple 1D graphical analysis can bring insights into the evolutionary dynamics at both  the RE and post-RE time scales.

Let us consider a population that is initially in the  state $x_0\in[0,1]$ and let us consider the evolution $x(t)$ given by the RE $\dot x=-x(1-x)V'(x)$. 
i) If $x_0$ is an equilibrium point of the RE, then $x(t)=x_0$ for all $t$;
ii) If $x_0$ is not an equilibrium point of the RE, then $x(t)\stackrel{t\to\infty}{\longrightarrow} x_1$, where $x_1$ is a stable equilibrium point of the RE (i.e., $V'(x_1)=0$ and $V''(x_1)>0$; for the sake of simplicity, we will ignore degenerate cases).

The post-RE phase, denoted (abusing notation) by $\tilde x(t)$, can be assumed to start at one of the equilibrium points; let us denote it by $\tilde x_0$ (equals to $x_0$ or $x_1$ in the previous notation).
If $\tilde x_0=0$ or $\tilde x_{0}=1$, then $\tilde x(t)=\tilde x(0)$ for all $t$. Let us now consider what happens when $\tilde x(0)\in(0,1)$.

By way of example, consider the family of potentials 
\[
V_\alpha(x)=5x\left(x-\frac{1}{2}\right)(x-1)+\alpha \frac{x}{16}\ .
\]
For $\alpha\in[-40,20]$, the RE flow is given by

\centerline{
	\begin{tikzpicture}
	%   \tikzstyle{every node}=[circle,2pt,draw=black,thick]
	\node[fill=black,circle,minimum size=1.5pt,draw=black,thick] (x0) at (0,0) {};
	\node[fill=white,circle,minimum size=2pt,draw=black,thick] (x1) at (2,0) {};
	\node[fill=black,circle,minimum size=2pt,draw=black,thick] (x2) at (4,0) {};
	\node[fill=white,circle,minimum size=2pt,draw=black,thick] (x3) at (6,0) {};
	\draw[thick,->] (x1) -- (x0);
	\draw[thick,->] (x1) -- (x2);
	\draw[thick,->] (x3) -- (x2); 
	\node[below of=x0,node distance=12pt] {$0$};
	\node[below of=x1,node distance=12pt] {$x_-$};
	\node[below of=x2,node distance=12pt] {$x_+$};
	\node[below of=x3,node distance=12pt] {$1$};
	\end{tikzpicture}
}
with $x_\pm(\alpha)=\frac{1}{2}\pm\frac{1}{60}\sqrt{300-15\alpha}$, where $x_-$ ($x_+$) is the unique interior local maximum (minimum, respect.) of $V$. We  define $V_+\bydef V(x_-)>V(x_+)\defby V_-$. 

For any initial condition $x(0)<x_-$ ($x(0)>x_-$), the RE is such that $x(t)\stackrel{t\to\infty}{\longrightarrow} 0$ ($x_+$, respect.). In the first case, and if $x(0)$ is smaller and not too close to $x_-$, we have $\tilde x(t)=0$ for all $t$ and therefore for any initial condition such that $x(0)$ is sufficiently smaller than $x_-$ then we expect extinction, i.e., $F_i\approx 0$.

\begin{figure*}
\includegraphics[height=0.28\textwidth]{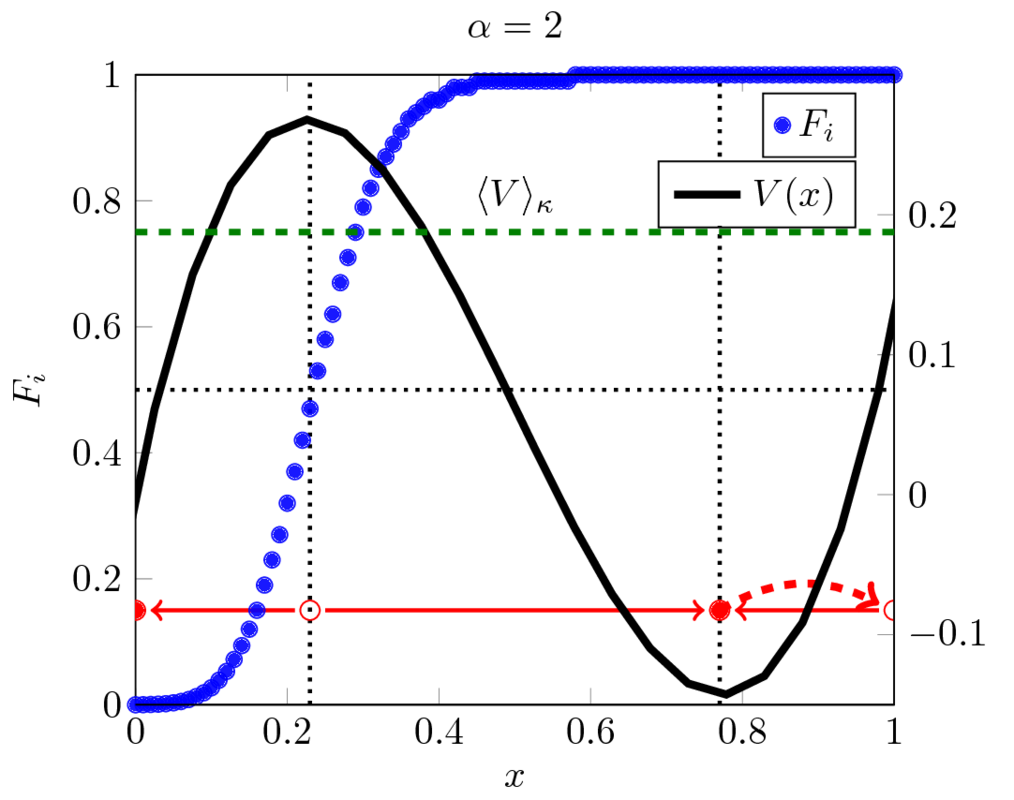}%
\includegraphics[height=0.28\textwidth]{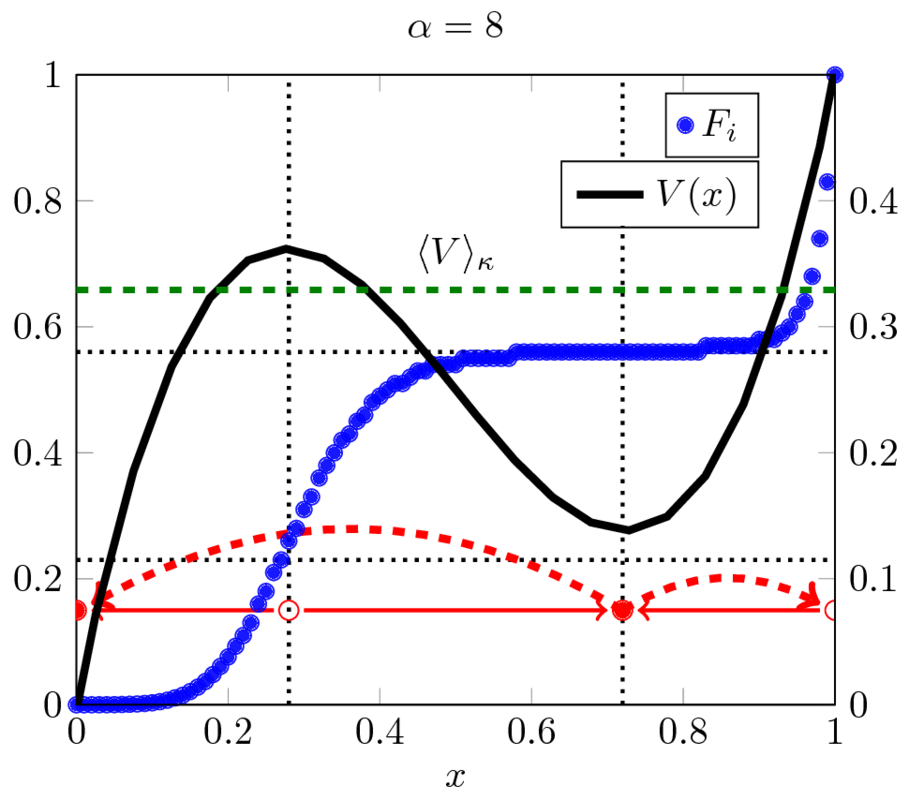}%
\includegraphics[height=0.28\textwidth]{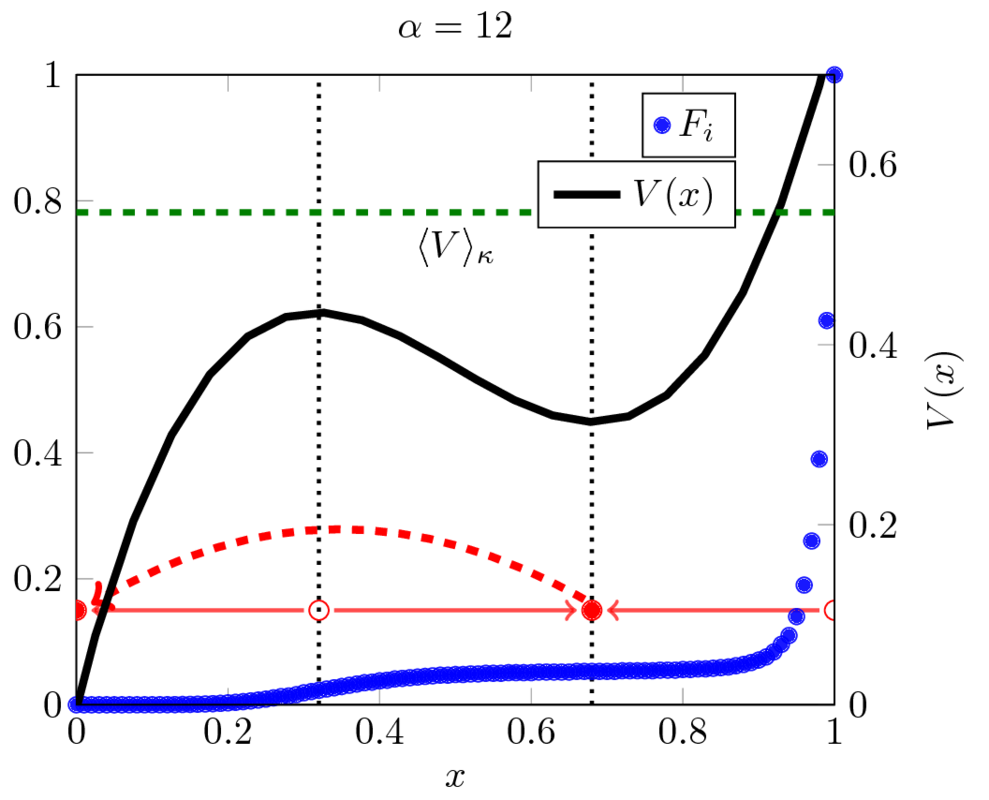}%
\caption{Potential $V_\alpha(x)=5x(x-1/2)(x-1)+\alpha x/16$ with $N=100$ and $\kappa=0.1$, indicated by thick black lines with values on the right axis. The green dotted line indicates the value of $\Vmean$. The fixation probability is indicated by thick blue dots with values on the left axis. For any initial condition $x(0)<x_-$, with $|x(0)-x_-|>\kappa$, $F_i\approx 0$. For $\alpha=2$, initial conditions above $x_-$ imply almost surely fixation. For the case $\alpha=12$, all initial conditions not too close to $x=1$ evolve towards extinction with probability close to one. 
		For, $\alpha=8$, $V_+\gtrsim\Vmean$ and therefore evolution is uncertain in the region $x(0)>x_-$. Note that the fixation is essentially independent of the initial state  of the system, in this region, as a consequence of the fact that the system converges initially to  $x_+$, in the RE time scale, and then, in the post-RE time-scale it reaches either  extinction or fixation. The RE and post-RE dynamics are indicated in continuous and dashed red lines, respect.  The potential axis scales were chosen differently for each graph to provide a better visualization. Note that the fixation $\varphi(x)$ increases only in the region such that $V(x)>\Vmean$, otherwise it is constant, consistently with Eq.~(\ref{eq:fix_prob_continous}) for fixed $\Vmean$ and small $\kappa$.}
	\label{fig:fourvaluesofalpha}
\end{figure*}

\begin{figure*}
	\centering
\includegraphics[height=0.28\textwidth]{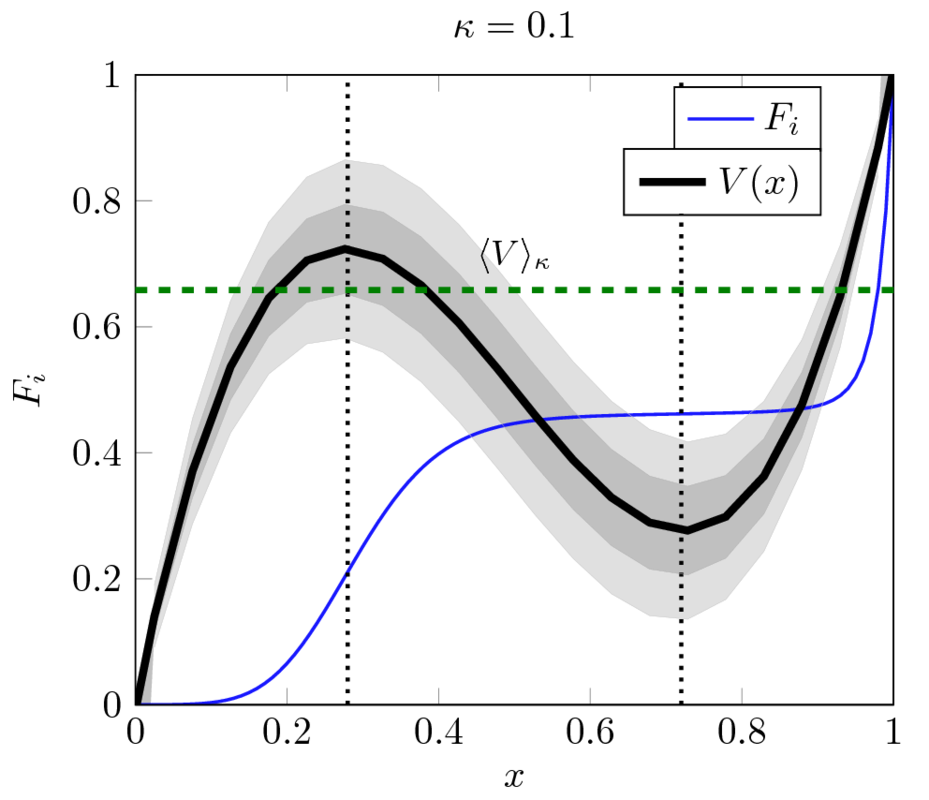}%
\includegraphics[height=0.28\textwidth]{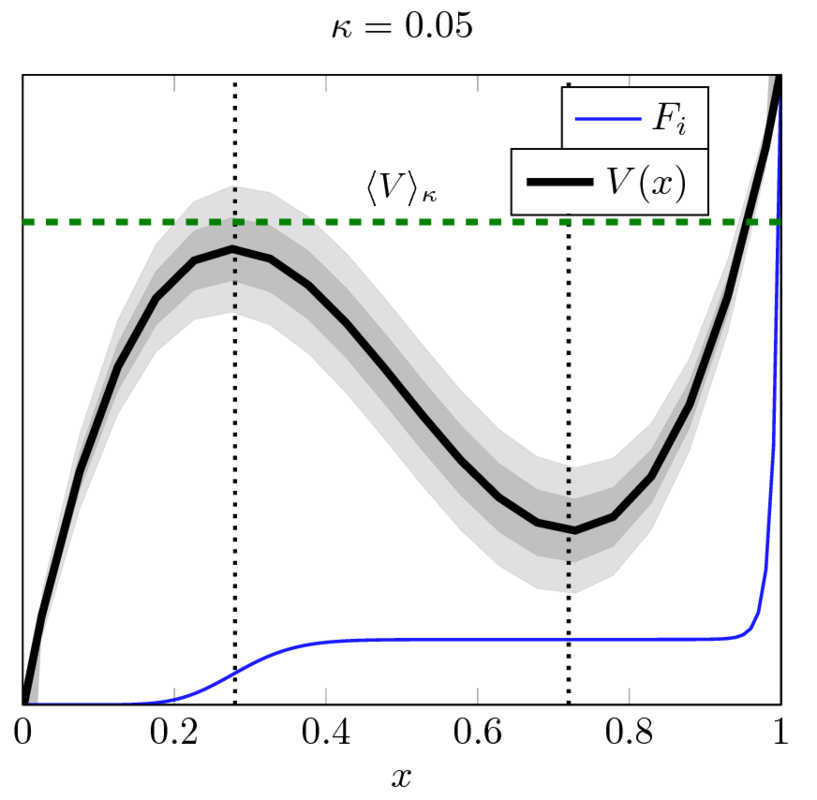}%
\includegraphics[height=0.28\textwidth]{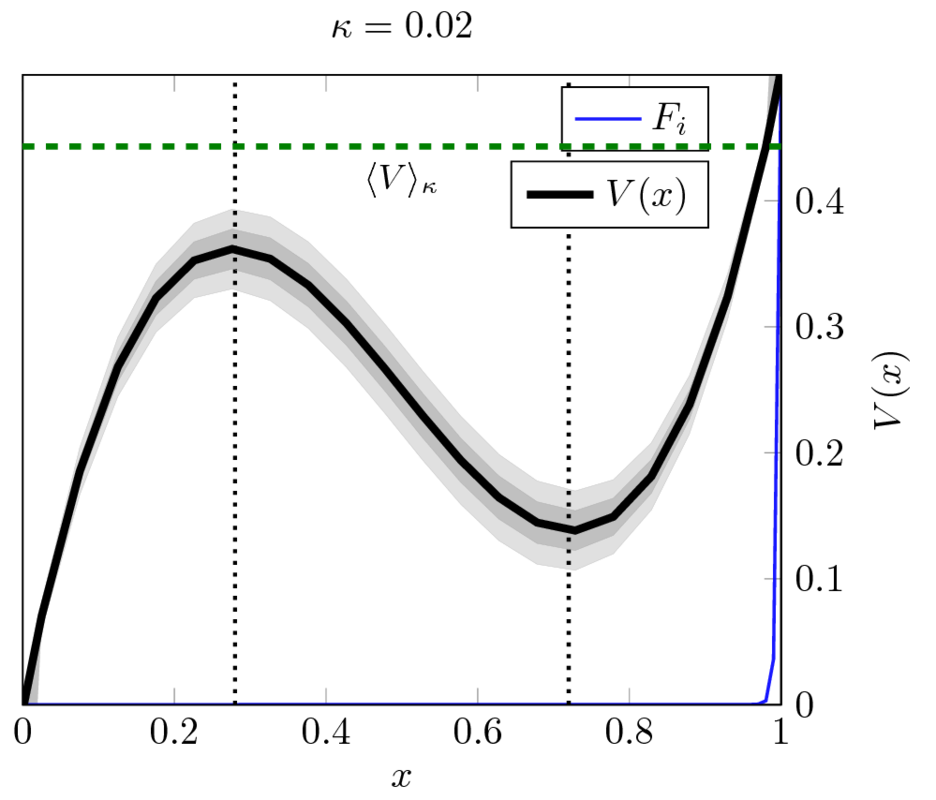}%
	\caption{Same potential as  of Fig.~\ref{fig:fourvaluesofalpha}, $\alpha=8$, $N=500$, but with $\kappa=0.1$ (left), $\kappa=0.05$ (centre) and $\kappa=0.02$ (right). 
		The fixation probability (blue continuous line) for the region attracted in the RE by $x_-$ decreases with the decrease of $\kappa$. The gray shaded zone indicates one half and one multiples of the standard deviations $\pm\sqrt{\kappa x(1-x)/2}$, the typical fluctuation of the WF process. The larger and darker is the gray region above $\Vmean$ at $x_-$, the larger is the probability to cross the local maximum at $x_-$ and be extinct.}
	\label{fig:threevaluesofkappa}
\end{figure*}

Now, looking at Figs.~\ref{fig:fourvaluesofalpha}~and~\ref{fig:threevaluesofkappa}, let us see what happens with the WF process if $\tilde x(0)=x_+$. 
Recall that we are assuming $N\kappa$ large enough such that the probability of selection for reproduction, given by equation~(\ref{eq:piweakselection}) is strictly positive, i.e., $p_\alpha(x)=x\left(1-\frac{1}{N\kappa}(1-x)\frac{\partial V_\alpha}{\partial x}\right)>0$ for all $x\in[0,1]$. 

The fixation pattern can be inferred by a simple graphical analysis. 
Note that if $V(1)>\Vmean>V_+$ ($V_+>\Vmean>V(1)$), then 
$\int_{0}^{x_+}\e^{\frac{2}{\kappa}(V(y)-\Vmean)}\rd y<x_+ 
\,\left(\int_{x_+}^1\e^{\frac{2}{\kappa}(V(y)-\Vmean)}\rd y<1-x_+, \text{respect.}\right)$, and therefore the necessary work for type \B (\A, respect.) to fixate is smaller  than in the neutral case; we expect, therefore, a small (large, respect.) fixation probability of type \A. Looking at the graphics, we see that if the current initial state is $x_+$, then the larger fixation probability will be associated to the boundary that can be reached with less energy. See Fig.~\ref{fig:fourvaluesofalpha} (cases $\alpha=2$ and $\alpha=12$).

The situation is  different when $\Vmean<\min\{V_+,V(1)\}$. In this case, the assumption $\kappa\ll1$ (i.e., the selection-driven regime) comes into full force in order to provide a better insight. Let $x_*(\alpha)\in\{x_-,1\}$ be the argument of the global maximum of $V_\alpha$ and $|V_+-V(1)|\gg\mathrm{O}\left(\kappa\right)$.  From Eq.~(\ref{eq:fix_prob_continous}), with $x=x_+$,
it follows from a straightforward application of Laplace's method \cite{Hinch1991} that, if $x_*=x_-$, then $\varphi(x_+)\stackrel{\kappa\to0}{\longrightarrow} 1$ and (once more) the system will typically converge to $x=1$ (fixation is almost certain); if $x_*=1$, then $\varphi(x_+)\approx\e^{\frac{2}{\kappa}(V(x_+)-V(1))}\stackrel{\kappa\to0}{\longrightarrow}0$, and extinction is almost certain. Note that the previous heuristics apply again; see Fig.~\ref{fig:threevaluesofkappa}. For small $\kappa$, but not too close to 0, with $\Vmean<\min\{V_+,V(1)\}$, the system seems to be ``undecided'', with comparable extinction and fixation probabilities --- see, once again, Fig.~\ref{fig:fourvaluesofalpha}, case $\alpha=8$.

\begin{figure}
 \includegraphics[width=0.95\textwidth]{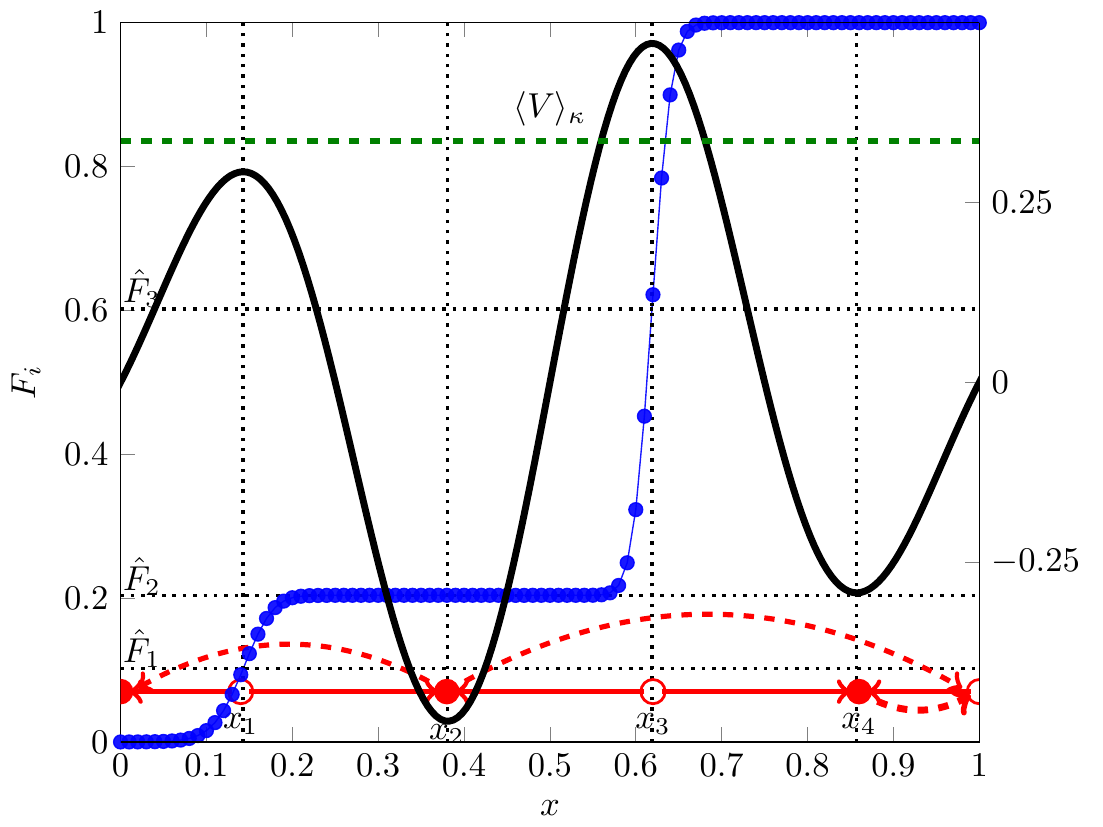}
\caption{The fitness potential is given by $V(x)=\frac{1}{2}\e^{-(2x-1)^2}\sin(4\pi x)$, $N=100$ and $\kappa=0.1$. The RE is given by the red line: $x=0$, $x_2$ and $x_4$ are attractors, while $x_1$, $x_3$ and $x=1$ are repellers. Therefore, except for $x(0)=x_1$ ,$x(0)=x_3$ or $x(0)=1$, the system will initially converge to $0$, $x_2$ or $x_4$. The first point is a stationary point of the WF process; the last one will be attracted to $x=1$, as $V(x_3)>\Vmean$. 
For $\tilde x(0)=x_2$, it is possible to go to the left or to the right, as $V(x_1)$, $V(x_3)$ are not much smaller than $\Vmean$ (with respect to $\kappa$; however, for smaller values of $\kappa$ it cannot reach the boundary $x=1$)
and it is also possible to go to the right (with probability $\hat F_2\approx 0.20$. If the initial condition is $x_1$, then it will go to the left with probability $1/2$, and therefore will be extinct, and to the right with probability $1/2$; in this case, it reaches fixation with probability $\hat F_{2}$; we conclude that $\hat F_1=\hat F_2/2$. A similar argument shows that $\hat F_3=(1+\hat F_2)/2$. Note that within each basin of attraction of the RE, the fixation probability is almost constant. This potential cannot be exactly generated by a $d$-player game, with $d$ finite, but a good approximation can be found with $d=6$, and pay-offs 
$\mathbf{A}=(
		5.56,  
		- 2.71,  
		1.46,   
		- 0.286,  
		0.270,  
		0.318)$
		and
		$\mathbf{B}=(
		7.13,  
		- 4.61,  
		3.04,  
		- 1.01,  
		0.360,  
		0.349)$, 
		where $A_k$ ($B_k$) is the pay-off of the \A (\B, respect) player when $k$ individuals play strategy \A. See~\cite{ChalubSouza_ArXiv2018} for a discussion about the technique used to obtain these values.}
	\label{fig:bizarre}
\end{figure}

In Fig.~\ref{fig:bizarre} we present  an example using the potential $V(x)=\frac{1}{2}\e^{-(2x-1)^2}\sin(4\pi x)$, where all the previous phenomena are present. This potential cannot be exactly obtained using evolutionary game theory with a finite number of players, but it can be well approximated by such a game --- cf. \cite{ChalubSouza_ArXiv2018}. 

Another important issue is  what happens when the initial condition is an unstable equilibrium of the RE --- cf. $x_-$ in Fig.~\ref{fig:fourvaluesofalpha} and $x_1,x_3$ in Fig.~\ref{fig:bizarre}. In this case, we expect the system to move away from equilibrium with probability $1/2$ to the left and $1/2$ to the right (if $x_+$ is far from the boundaries, the equilibrium distribution can be approximated by a Gaussian centred in $x_+$). If $V_+<\Vmean<V(1)$, then with probability $1/2$ it converges directly to 0 and with probability $1/2$ it goes to $x_+$ before going back to 0. Therefore, $F_{Nx_-}\approx 0$. However, if $V_+>\Vmean>V(1)$, again with probability $1/2$ the state of the system converges directly to 0, but with probability $1/2$ it goes first to $x_+$ and then to 1. Therefore, $F_{Nx_-}\approx 1/2$. . For the degenerate cases (i.e., when $V_+\approx V(1)$), we might consider small perturbations such that $V(x_-)>V(1)$ and $V(x_-)<V(1)$ to see that any value for the fixation probability is possible for $x\in[x_-,1]$. See~\cite{ChalubSouza16a} for further details.

\begin{figure}
	\includegraphics[width=0.99\textwidth]{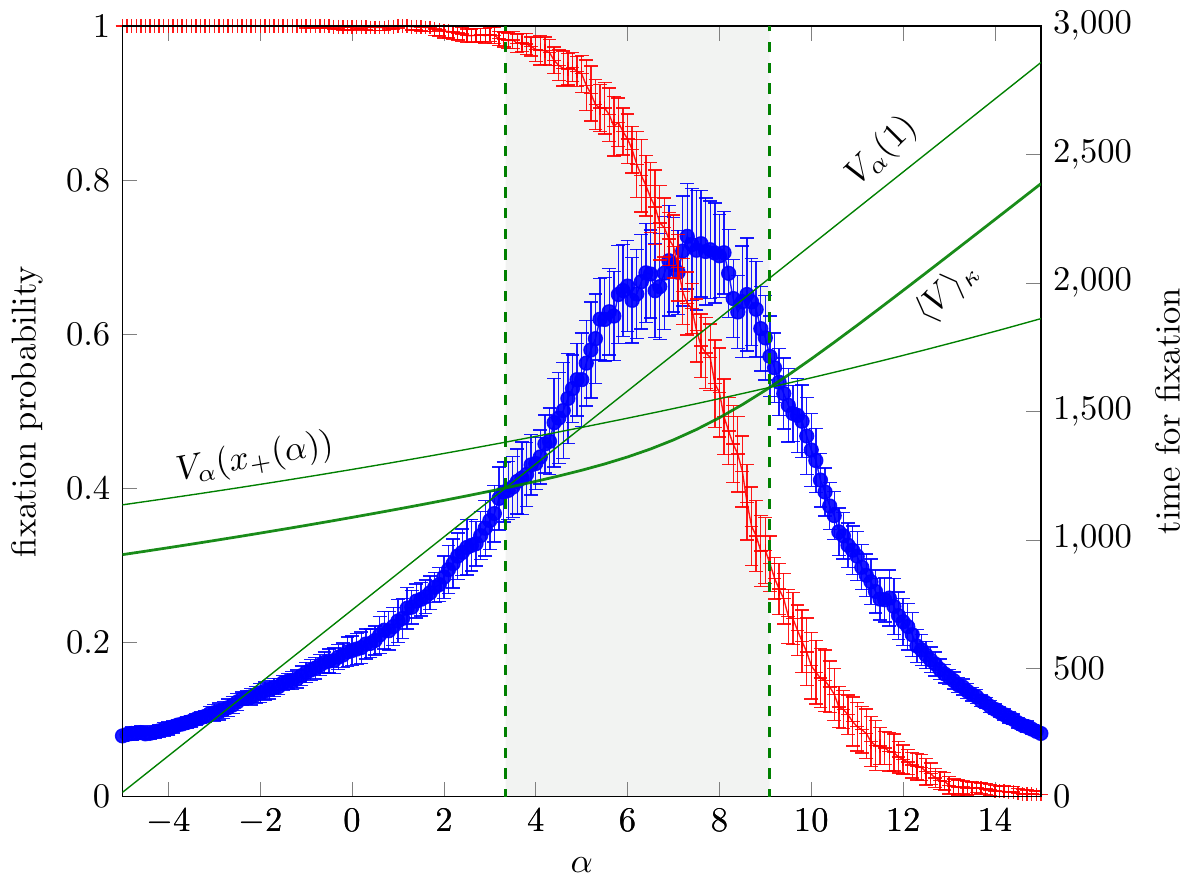}
	\caption{Fixation probability for the focal type (blue dots, left scale) and fixation times for any type (red crosses, right scale) for the fitness potential $V_\alpha$ and initial condition at $x_-(\alpha)$. We used $N=100$ and $\kappa=0.1$. Note that expected time for fixation is much larger when it is is uncertain, meaning that the system has no preferred direction when it starts on the stable equilibrium of the RE. The thin continuous green lines indicate $V_+$ and $V(1)$ (no scales), while the thick green line indicates $\Vmean$. Note that the time for fixation is much larger when $\Vmean$ is smaller than both $V_+$ and $V(1)$ (delimited by horizontal dashed green lines) indicating that the region easily accessible from $x_+$ does not include any absorbing state, and therefore the system takes much longer to fixate or extinct. Note that this is the only figure in this work where we resorted to explicit numerical simulation of the WF process, and not to numerical solution of linear equations — this was done only to illustrate the use of both techniques.}
	\label{fig:times}
\end{figure}

Whenever the system seems to be ``undecided'' in which way to go, the expected time until the final state is reached is significantly higher than 
otherwise. Fig.~\ref{fig:times} shows the expected fixation time for any type under the potential $V_\alpha$, and  for initial condition $x_-(\alpha)$.
Assume $\tilde x(0)=x_+$ and consider a maximal interval $I\ni x_+$ such that $V(x)<\Vmean$ for all $x\in I$. If $\alpha<\alpha_-\approx 3.3$ then  $1\in I$, but $0\not\in I$ and therefore fixation is almost certain. If $\alpha>\alpha_+\approx 9.1$ then $1\not\in I$ and $0\in I$, therefore extinction is almost certain. However, for $\alpha_-<\alpha<\alpha_+$, $0,1\not\in I$, and fixation or extinction occurs with comparable likelihood. If we understand $I$ as the interval of states that can be typically achieved by stochastic fluctuations, the fact that both stationary states are outside these intervals implies that the time necessary to reach fixation or extinction will be larger.   Therefore, for $\alpha\in(\alpha_-,\alpha_+)$ the state $x_+(\alpha)$ presents an stability for the WF process much similar to what happens in the RE. In this sense, we propose a new definition of evolutionary stability for finite populations, with focus in the effective population size: 

\begin{definition}
$x_*\in(0,1)$ is an $\ess_\kappa$ if a) $x_*$ is a stable equilibrium of the RE; and b) $V(x_*)<\Vmean$ and c)
the maximal interval $I\ni x_*$ such that $V(x)<\Vmean$ for any $x\in I$ is such that $0,1\not\in I$. If $x_*=0$ or 1, the third condition is replaced by saying that $1-x_*\not\in I$.
\end{definition}

If $x=0$ (or $x=1$), these conditions imply the continuous versions given in \cite{ChalubSouza16a} of the standard definition of $\ess_N$ for monomorphic states~\cite{Nowak:06}. 
Furthermore, if $x$ is an $\ess_\kappa$, then $\varphi(x)$ has a plateau around $x$, and the value of $\varphi$ in this plateau is not close to zero or one. 

\begin{figure}
\centering
\includegraphics[height=0.28\textwidth]{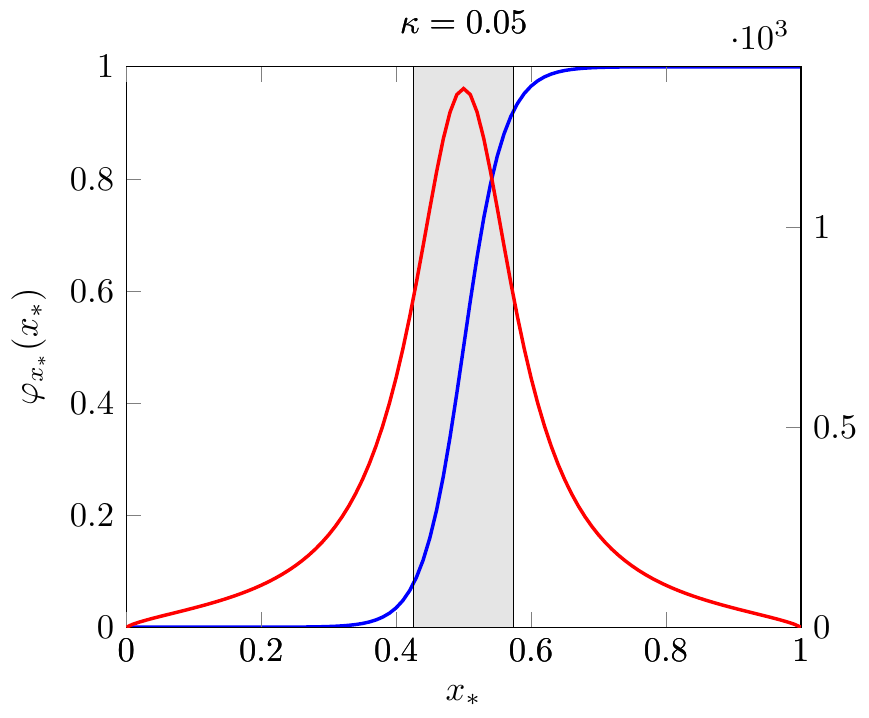}
\includegraphics[height=0.28\textwidth]{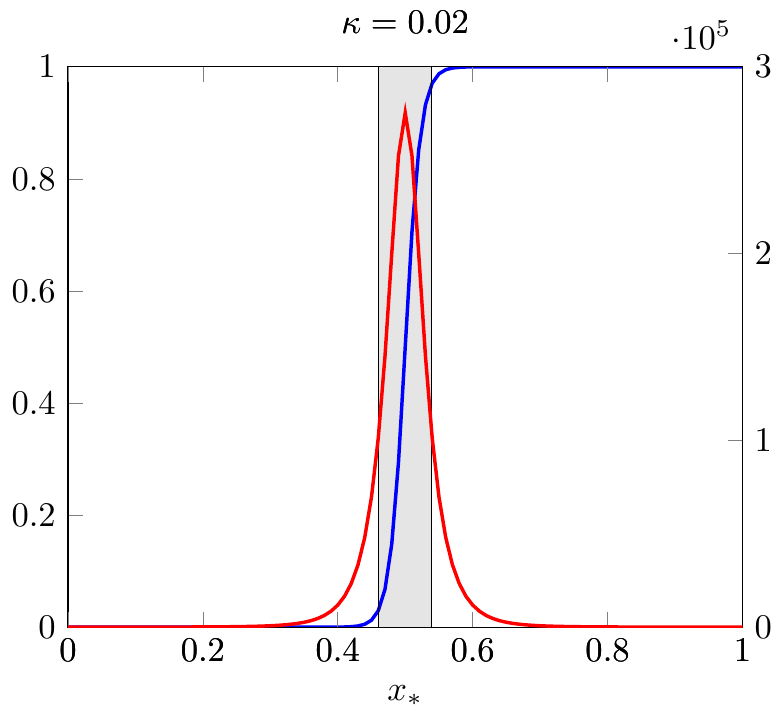}
\includegraphics[height=0.28\textwidth]{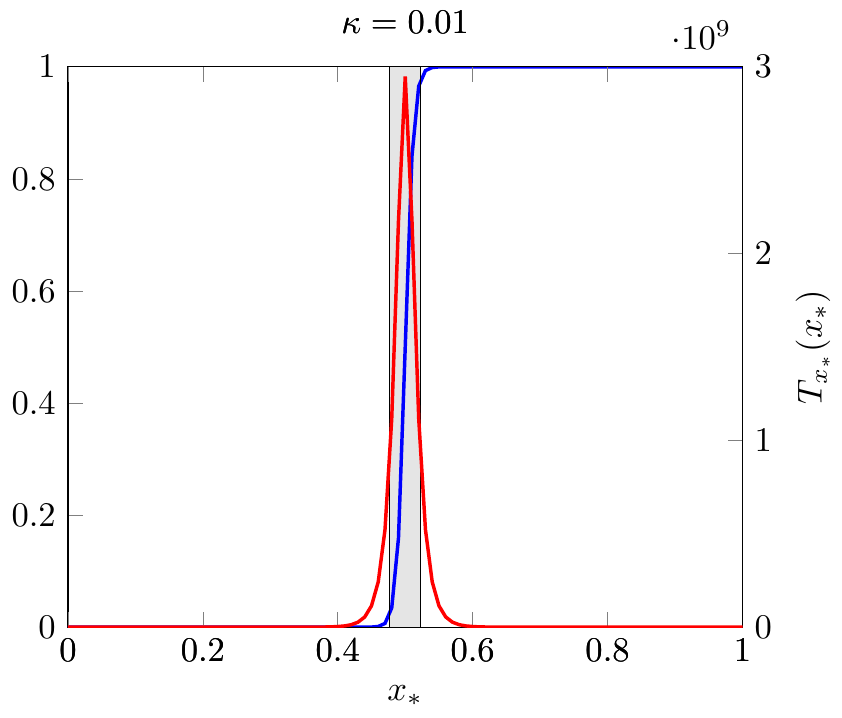}
	 \caption{Fixation probability (blue, left scale) $\varphi_{x_*}(x_*)$ and time for fixation (red; diferent scales, marked on the right scale, for different figures) $T_{x_*}(x_*)$ for the potential $V_{x_*}(x)=-xx_*+x^2/2$, initial condition $x=x_*$ (minimum of the potential) and different values of $\kappa$. To the left (right) of the gray area, we have that $V_{x_*}(0)<\langle V_{x_*}\rangle_\kappa<V_{x_*}(1)$ ($V_{x_*}(0)>\langle V_{x_*}\rangle_\kappa>V_{x_*}(1)$, respect.) and therefore extinction (fixation, respect.) is almost certain. In the gray area, $\langle V_{x_*}\rangle_\kappa<\min\{V_{x_*}(0),V_{x_*}(1)\}$, i.e., $x_*$ is an $\ess_\kappa$ and the final outcome is uncertain and the system takes significantly more time to reach the final state. Note that the scales for the time for fixation vary in several orders of magnitude. In particular, the entire family of potentials $V_{x_*}$, $x_*\in[0,1]$ is such that only $x_*=\sfrac{1}{2}$ is an $\ess_\kappa$ for all values of $\kappa$.
	 }
	 \label{fig:concave}%{fig:ess_HD}
\end{figure}

Note that $\lim_{\kappa\to0}\Vmean\uparrow\max_{x\in[0,1]}V(x)$.%, from below. 
Therefore, any potential $V$ with an unique global maximum $V_{+}$ has no $\ess_\kappa$ for $\kappa$ small enough. See Fig.~\ref{fig:concave} for a family of concave quadratic potentials such that the associated replicator dynamics have a mixed $\ess$ (coexistence in game theory parlance),
whereas the corresponding WF dynamics has an $\ess_\kappa$ only for some of the family members -- in this case, the fixation work for both 
types, when the population is in the interior minimum of the potential, is comparable within $\mathcal{O}(\kappa)$.

	In Fig.~\ref{fig:concave}, we examine the interplay between   fixation probability, expected time for fixation and the  $\ess_\kappa$ equilibrium range for 
	two-player coexistence games. We recall that, as shown in \cite{ChalubSouza16a}, the fixation probability in such games for small $\kappa$ and different values of the ESS equilibrium is mostly of dominance type  except for an order $\kappa$ region around $x^*=\sfrac{1}{2}$ --- this was termed the  $\sfrac{1}{2}$-law. The ESS$_\kappa$ range is essentially the same, and inside the region the expected fixation times increase significantly as the equilibrium approaches $\sfrac{1}{2}$. Notice also that there are antagonistic effects for a fixed population size: while the $\ess_\kappa$ regions increases with larger  $\kappa$, the expected fixation time increases with smaller values of $\kappa$.

\section{Further Developments and Discussion}

The theory developed so far provides a reinterpretation of two-types evolutionary models, which includes many different models including the Moran and Wright-Fisher processes, the Kimura equation and the Replicator Dynamics. It is also worthwhile noticing that, while we have not attempted to provide a full mathematical analysis, much of this reinterpretation is based on available rigorous mathematical theory \cite{ChalubSouza16a}. Naturally, one might expect that such a reinterpretation program can be carried over to more general situations. In order to assess the potential of such generalizations, we will use its central ideas for two different numerical studies: in Subsec~\ref{sec:3T} we will analyse three-types (3T) evolution, while in Subsec.~\ref{sec:mutations} we will study 2T models with mutations.

\subsection{Three types evolution}
\label{sec:3T}

\begin{figure*}[p]
\includegraphics[width=0.32\textwidth]{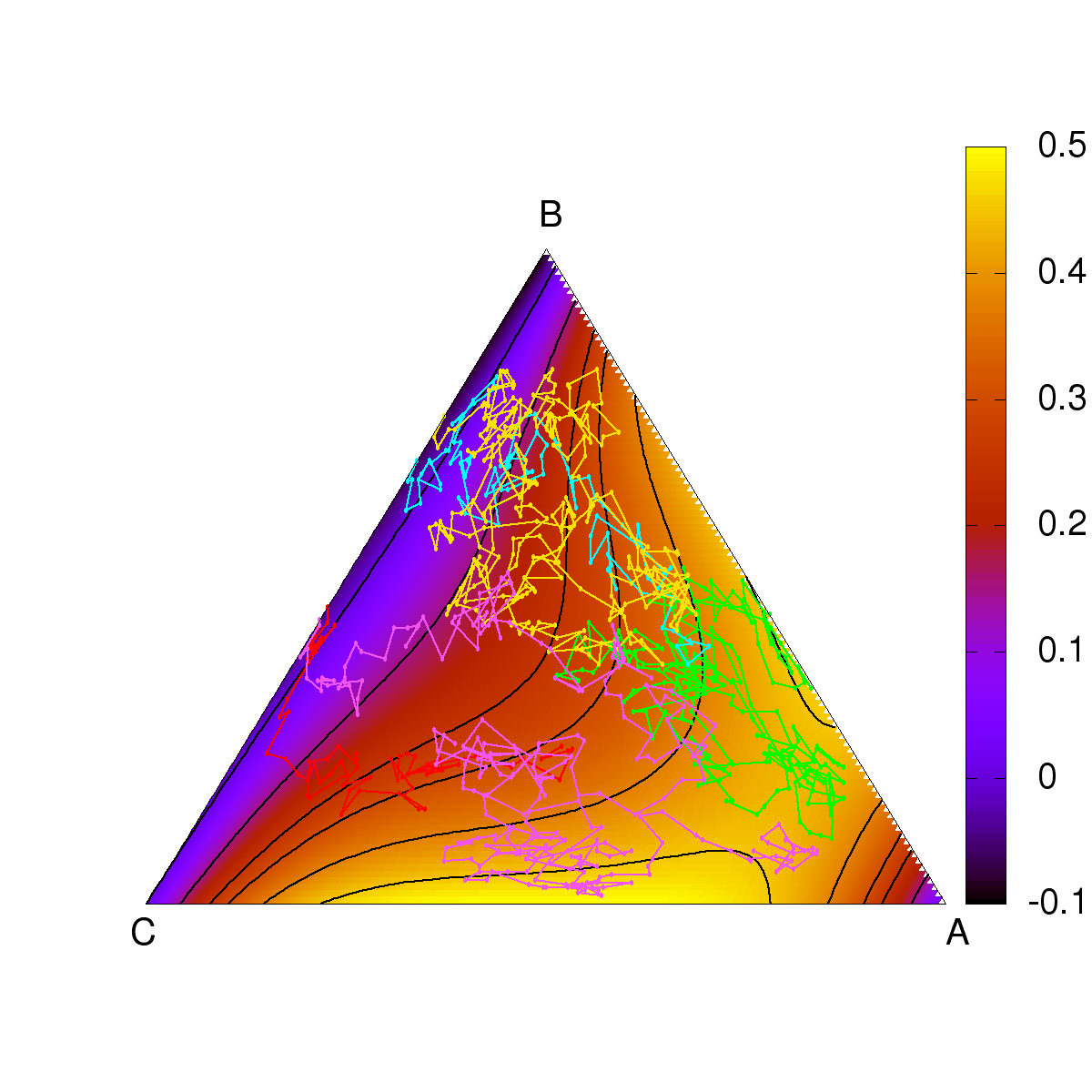}
\includegraphics[width=0.32\textwidth]{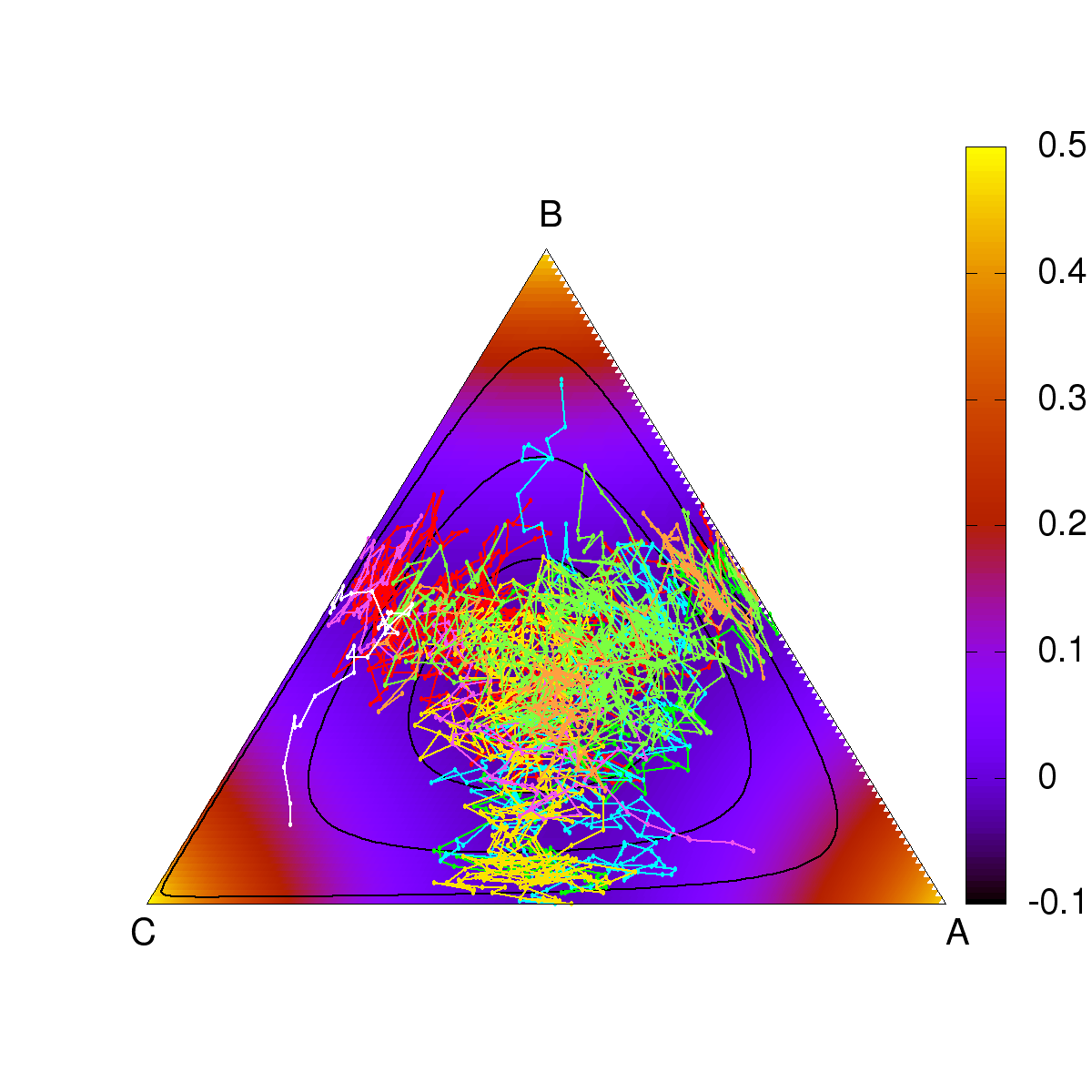}
\includegraphics[width=0.32\textwidth]{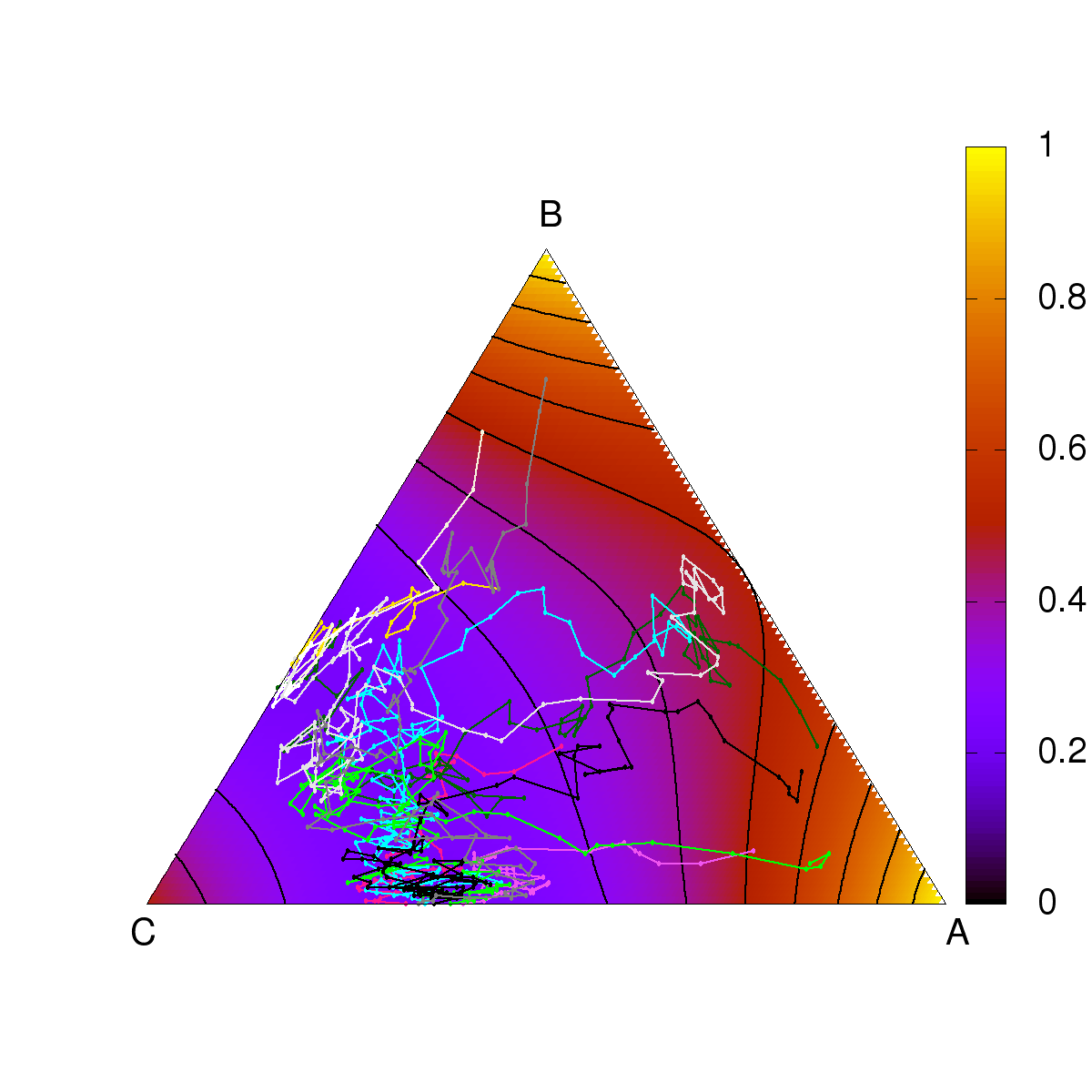}
\includegraphics[width=0.32\textwidth]{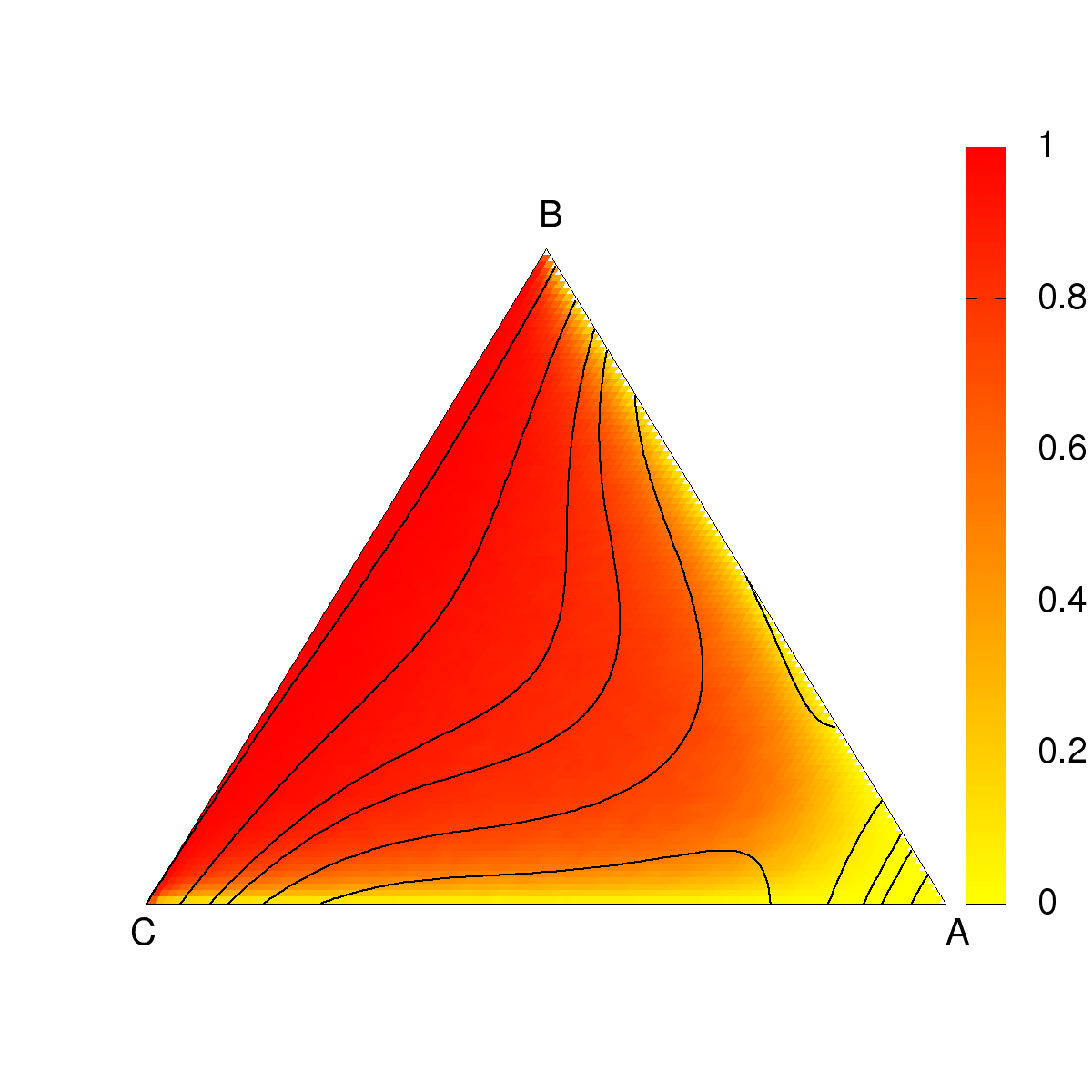}
 \includegraphics[width=0.32\textwidth]{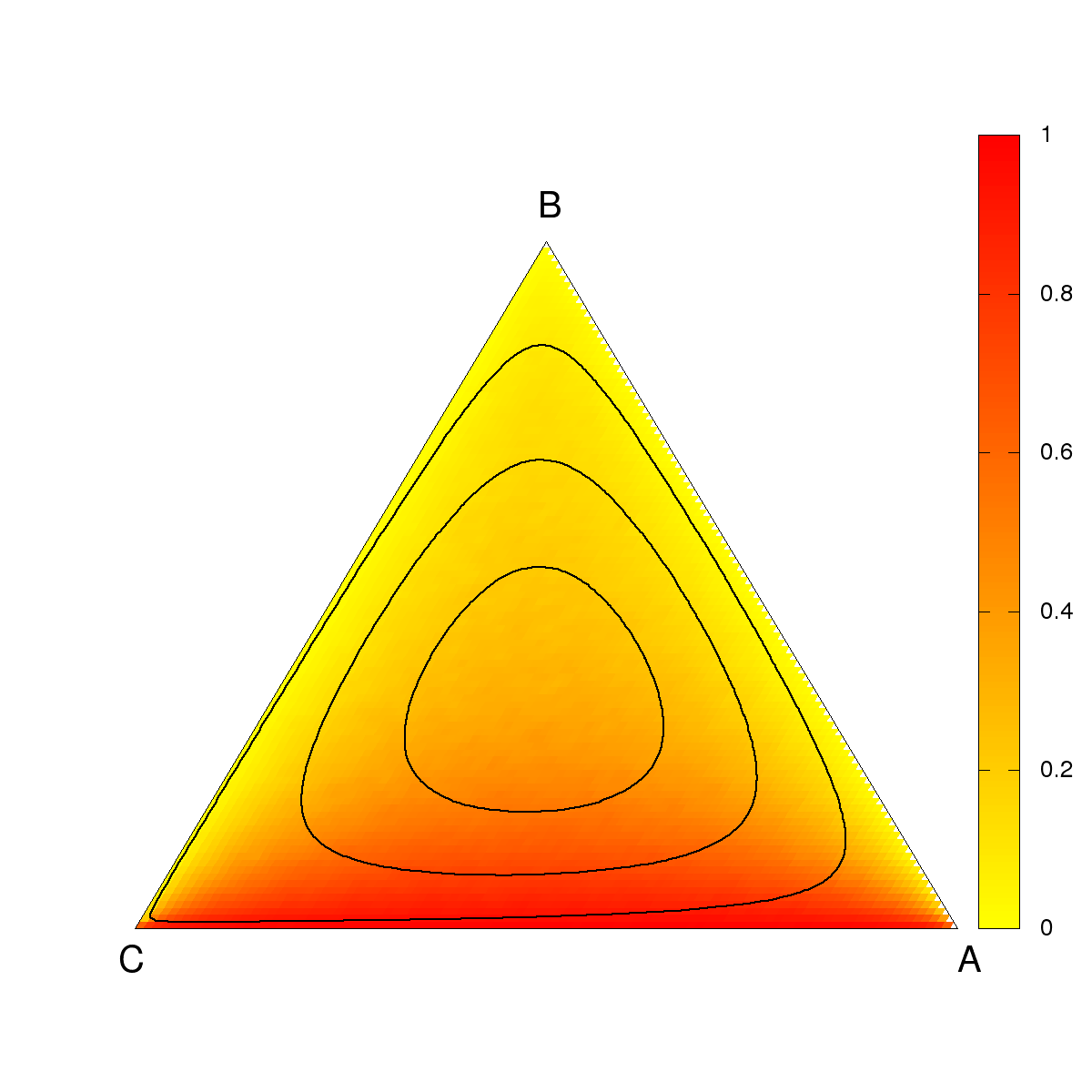}
 \includegraphics[width=0.32\textwidth]{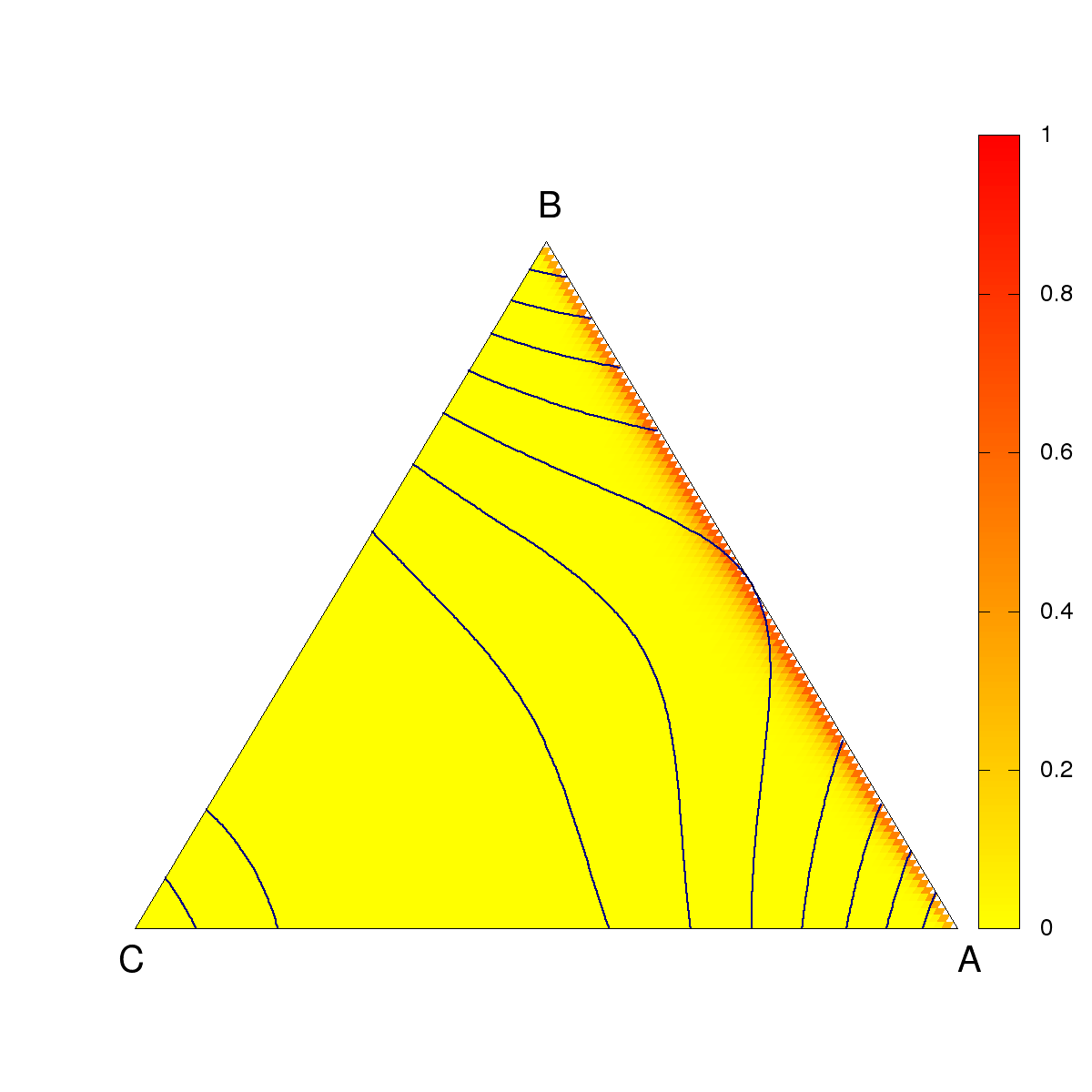}
 \caption{First column: Above: Evolution of the WF process with $\kappa=0.1$, $N=250$ and $V(x,y,z)=4x(1-x)((x-1/2)^2+(y-1/2)^2+(z-1/2)^2)-y/10$. Note that typically the first extinction is of the \A type, representing the edge such that the minimum of the potential is minimum ($\min V$ is $-0.1$, $0.25$ and $0.208$ for the  faces opposing vertexes representing monomorphic populations of \A, \B and \C type, respectively). Below: probability that the first extinction is of  type \A, as function of the initial condition. The level curves, in both cases, are the level curves of the potential $V$. Second column: Same $N$ and $\kappa=0.02$, $V(x,y,z)=4(x-1/2)(y-1/2)(z-1/2)-x/30-y/50$. In each face, the minimum is attained close to the centre point, with values $-0.010$, $-0.017$, and $-0.027$ for the faces opposing vertexes representing monomorphic populations of \A, \B and \C type, respectively, always below $\Vmean$.  In particular, we do not expect significant  differences among the types as far as the first extinction is concerned (see figure below, for the probability that type \A is the first to be extinct). Last column: same $N$ and $\kappa$ and $V(x,y,z)=4(x-1/2)(y-1/2)(z-1/2)+(x+y)/2$, i.e., such that $V(0,0,1)\le V(x,1-x,0)$ for any $x$ (with equality if and only if $x=1/2$). We expect a low probability that the first extinction is of type \C (see below).}
 \label{fig:3T_first_extinction}
\end{figure*}

Let us consider a potential $V:\R_+^n\to\R$, and define fitness functions $\psi^{(i)}=-\partial_iV$. The presence of each type is given by $x_i$ and the dynamics is confined to the $n-1$ dimensional simplex 
$S^{n-1}=\{\bx\in\R^n|x_i\ge 0, \sum_ix_i=1\}$. The RE is given by $\dot x_i=-x_i\left(\partial_iV-\overline{\nabla V}\right)$, where $\overline{\nabla V}=\sum_ix_i\partial_iV$. 
We will now show how our developed formalism for 2 types, $n=2$, can be consistently extended for three types, $n=3$.
\[
 p_i(\bx)=x_i\left[1-\frac{1}{\kappa N}\left[\partial_iV-\overline{\nabla V}\right]\right]\ .
\]
Usually one of the types will be extinct first, and in Fig.~\ref{fig:3T_first_extinction} we plot numerical simulations of the WF process for certain fitness potentials and also present some numerics on the probability that a given type is the first to be extinct.

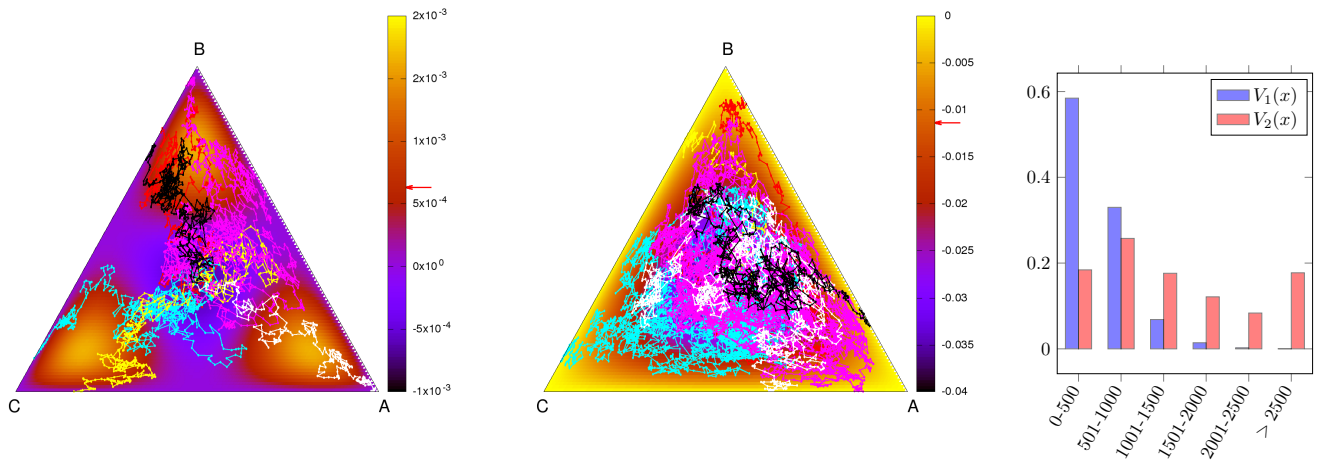
\begin{figure*}[p]
\includegraphics[width=0.38\textwidth]{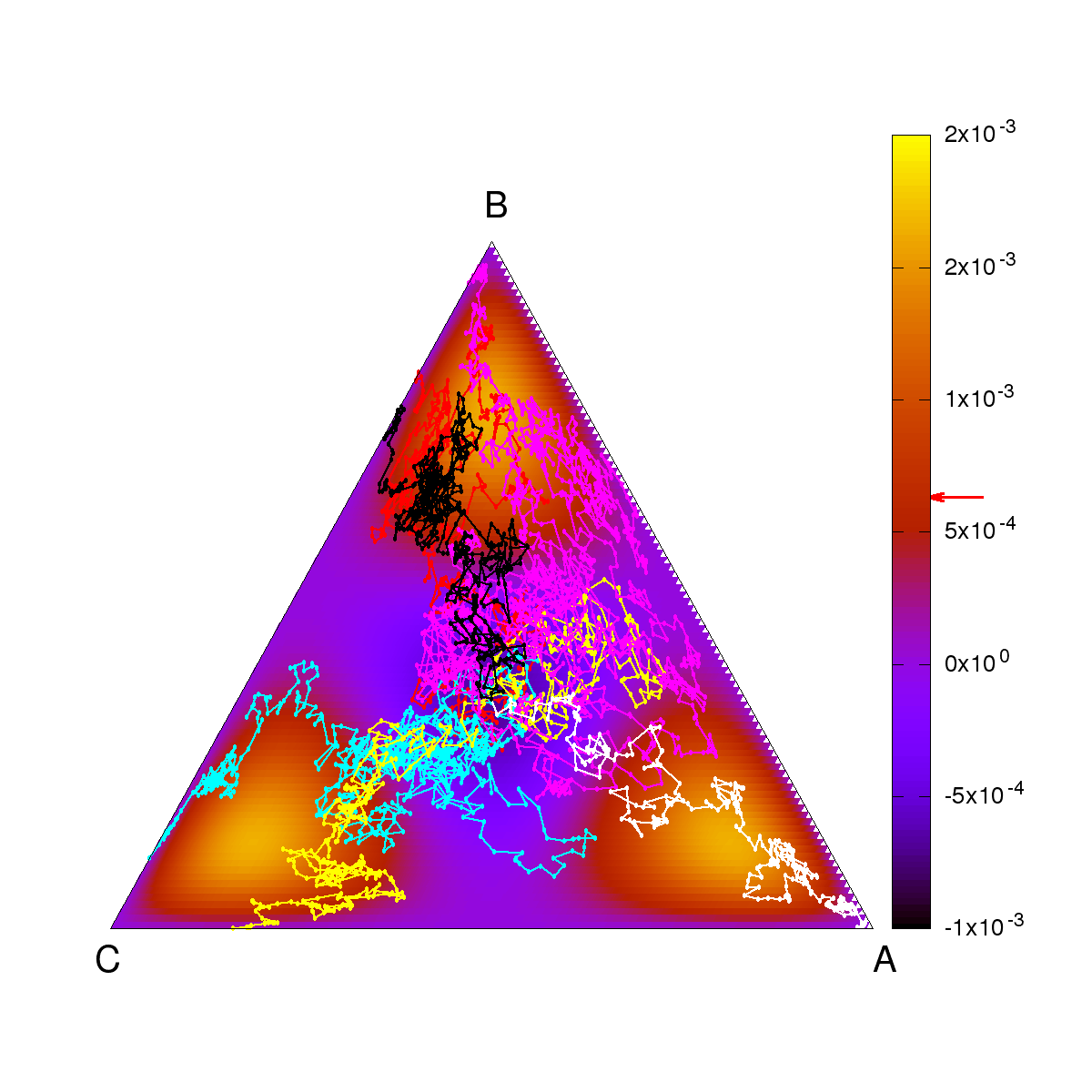}
\includegraphics[width=0.38\textwidth]{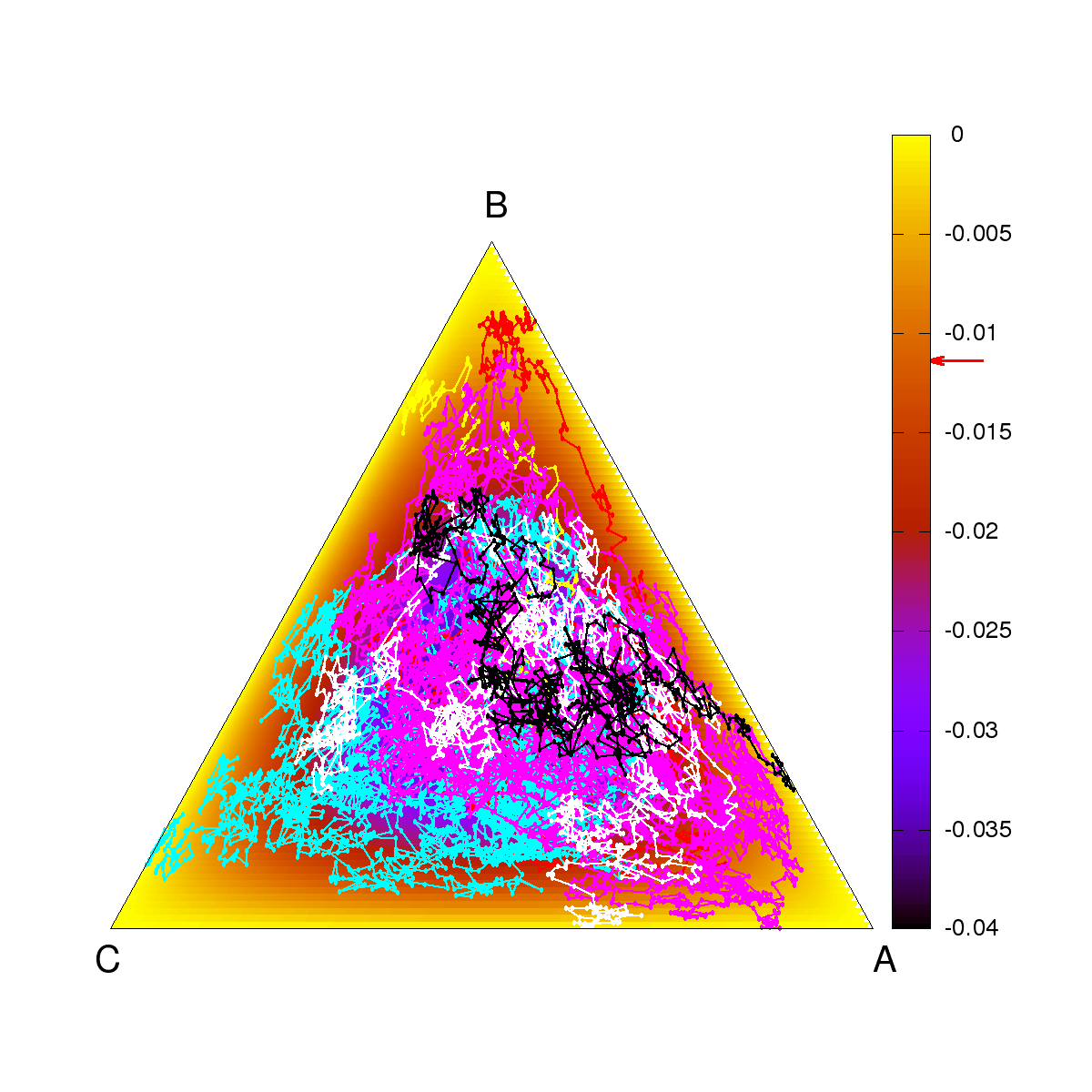}
\begin{tikzpicture}[scale=0.7]
\begin{axis}[ybar=0.5pt,x=0.8cm,bar width=7pt,ybar=0pt, symbolic x coords={0-500,501-1000,1001-1500,1501-2000,2001-2500,$>2500$},xtick=data,xticklabel style={rotate=60,anchor=east},legend style={ area legend}]
\addplot[opacity=0.5,fill=blue,area legend]
coordinates {(0-500,0.58461)
(501-1000,0.32999)
(1001-1500,0.06833)
(1501-2000,0.01417)
(2001-2500,0.00244)
($>2500$,0.00046)};
\addplot[opacity=0.5,fill=red,area legend]
coordinates {(0-500,0.18413)
(501-1000,0.25755)
(1001-1500,0.17624)
(1501-2000,0.12126)
(2001-2500,0.08352)
($>2500$,0.1773)};
\legend{$V_1(x)$, $V_2(x)$}
\end{axis}
\end{tikzpicture}

\caption{Left: $V_1(x,y,z)=4(x-1/2)(y-1/2)(z-1/2)xyz$. Centre: $V_2(x,y,z)=-xyz$. In both cases, we have a symmetric potential in $x$, $y$ and $z$, with a minimum at $\bx=\left(\sfrac{1}{3},\sfrac{1}{3},\sfrac{1}{3}\right)$. We consider several simulations with $N=900$ and $\kappa=0.02$ starting from $\bx$. Note that for $V_1$ the faces can be reached with potential energy less that $\Vmean$, while for $V_2$ this is not possible. See the escales, where $\Vmean\approx  4\times 10^{-4}$, and $-0.015$ for $V_1$ and $V_2$, respectively are marked in red). Right: probability that the first extinction happens in certain ranges of the number of generations, for potentials $V_1$ (blue) and $V_2$ (red), $N=900$, $\kappa=0.02$ and initial condition at the centre (only interior local minimum in both cases). Note that it is clear that the $V_2$ cases takes significantly more time until the first extinction, consistently to the idea that $\bx=\left(\sfrac{1}{3},\sfrac{1}{3},\sfrac{1}{3}\right)$ is an $\ess_\kappa$ only in the second case. }
\label{fig:3T_esskappaA}
\end{figure*}

\begin{figure*}[p]
\includegraphics[width=0.48\textwidth]{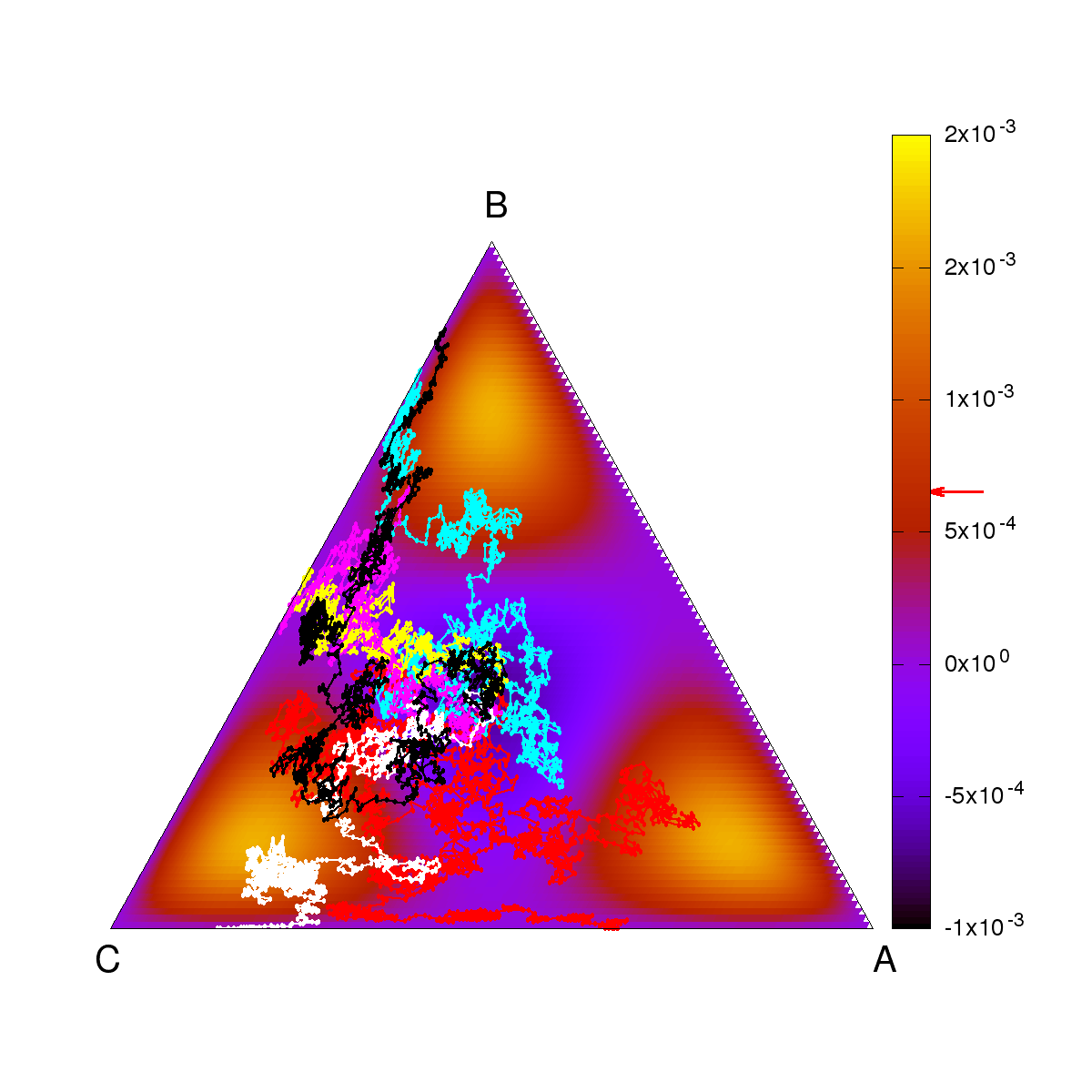}
\includegraphics[width=0.48\textwidth]{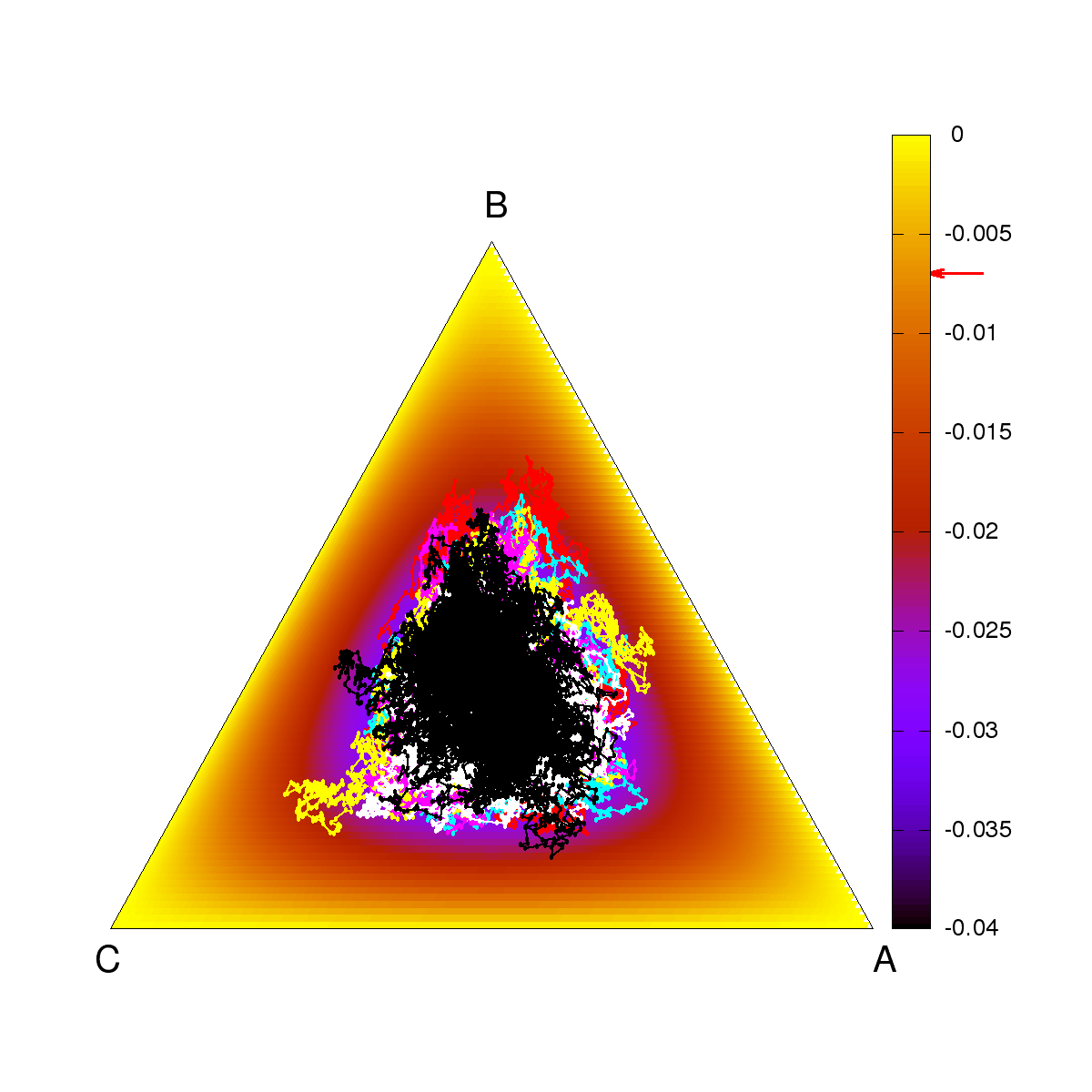}

\caption{ 
Same potentials as Fig.~\ref{fig:3T_esskappaA}, but with $N=3000$ and $\kappa=0.005$ starting from $\bx=\left(\sfrac{1}{3},\sfrac{1}{3},\sfrac{1}{3}\right)$ for up to 10 000 generation of the WF process or until one type is extinct. Note that for $V_1$ the faces can be reached with potential energy less that $\Vmean$, while for $V_2$ this is not possible. See the escales, where $\Vmean\approx  4.66\times 10^{-4}$, and $-0.00577$ for $V_1$ and $V_2$, respectively are marked in red). For the $V_1$ case, the first extinction occurs in $1804\pm1131$ generations, while in 1000 simulations of the $V_2$, no extinction occurred in less than 3000 generations, consistently to the idea that $\bx=\left(\sfrac{1}{3},\sfrac{1}{3},\sfrac{1}{3}\right)$ is an  $\ess_\kappa$ only in the second case.
}
\label{fig:3T_esskappaB}
\end{figure*}

Note that for $\kappa$ small enough, and assuming $\bx\in\mathrm{int}\, S^{n-1}$,  typically the first type to be extinct is among the types whose vertex is opposite to the faces that can be reached through a path 
$\gamma:[0,T]\to S^{n-1}$ such that $\gamma(0)=\bx$ and $\gamma(t)<\Vmean$ for all $t$. Furthermore, the typical trajectory of the WF process will follow valleys of $V$.
If no face can be reached in such way, then the typical time for the first extinction will be substantially larger and the interior minimum of $V$ is an $\ess_{\kappa}$. See~Fig.~\ref{fig:3T_esskappaA} and Fig.~\ref{fig:3T_esskappaB}.

These results suggest the following definition.

\begin{definition}
We say that $\bx_*\in\mathrm{int}\,S^{n-1}$ is an $\ess_\kappa$ if a) $\bx_*$ is a stable equilibrium of the RE; b) $V(\bx_*)<\Vmean$ and c) the maximal subset of $S^{n-1}$ $I\ni\bx_*$ with $V(\bx)<\Vmean$ for all $\bx\in I$ is such that $I\cap\partial S^{n-1}=\emptyset$. If $\bx_*\in\partial S^{n-1}$, we replace the third condition by $I\cap\partial S^{n-1}\subset\{\bx\in S^{n-1}|x_i=0,\forall i\not\in\mathop{\mathrm{supp}}\bx_*\}$, where $\mathop{\mathrm{supp}}\bx\bydef\{i|x_i>0\}$ is the \emph{support} of $\bx$. 
\end{definition}
 Here, a short comment is necessary. If $\bx_*\in\partial S^{n-1}$, then the $\ess_\kappa$ definition requires that the system will take a long time to escape from the neighbourhood of $\bx_*$ in case it is invaded by a small set of strategists, even if the invaders play strategies not in the support of $\bx_*$. Therefore, $\bx_*$ may be an $\ess_\kappa$ of the sub-game defined by strategies in $\mathop{\mathrm{supp}} \bx_*$, even if it not an $\ess_\kappa$ of the original game.

\subsection{Adding mutations}
\label{sec:mutations}

So far, we have studied only the class of what could be called \emph{conservative evolutionary dynamics}, also known as \emph{gradient systems}~\cite{HofbauerSigmund}. The first non-conservative example to be considered is a two-types system to which small mutation rates $\epsilon_{1,2}>0$, indicating rate of changes $\A\to\B$ and $\B\to\A$, respectiveley, are included. Let $\phi_1,\phi_2:[0,1]\to\R$ be the fitnesses of types \A and \B, respectively. The selection-mutator equation is given by~\cite{HofbauerSigmund}
\[
 \dot x=x(1-x)\left[f_1(x)-f_2(x)-\left(\frac{\epsilon_2}{1-x}-\frac{\epsilon_1}{x}\right)\right]=-x(1-x)V_\eff'(x)\ ,
\]
where
\[
V_\eff(x)=V(x)-\epsilon_1\log x-\epsilon_2\log(1-x)\ .
\]
Therefore, whenever $V$ is well defined in the interval $[0,1]$, the \emph{effective potencial} $V_\eff$ will present a logarithmic divergence close to both boundaries, i.e., $\lim_{x\to0,1}V_\eff(x)=+\infty$. This represents an effective repulsive force close to the boundary that prevents the system to reach the monomorphic state. In the interior of the domain its influence will be $\mathcal{O}(\epsilon_1,\epsilon_2)$. However, as will shortly see, $V_\eff$ will provide heuristic information in the dynamics of the system.

It is clear that 
\[
\left\langle V_{\eff}\right\rangle_k=\frac{\kappa}{2}\log\int_0^1\e^{2V(y)/\kappa}y^{-2\epsilon_1/\kappa}(1-y)^{-2\epsilon_2/\kappa}\rd y
\]
is finite if and only if $\epsilon_{1,2}<\kappa/2$. Therefore, if the mutation rate is sufficiently small (when compared to the genetic drift), then $\left\langle V_{\eff}\right\rangle_k$ will be finite and our theory is divergence-free. In this case, one expects a small impact on the Markov chain  dynamics, and indeed~\cite{fudenberg2006imitation} provides a simple algorithm to compute the limit  distribution for vanishing mutation rates. In \cite{wu2012small}, additional bounds on the mutation rate were derived, so that the invariant distribution is close, in the total variation norm, to the quasi-stationary distribution when there are no mutations. More recently, \cite{Vasconcelos_et_al_PRL2017} provided an enhanced hierarchical  approximation method for invariant distributions. From now on, following~\cite{Vasconcelos_et_al_PRL2017}, we call this regime the small mutation regime (SMR).   In this case,  we have that 
\[
 \varphi^{(\epsilon_1,\epsilon_2)}(x)=\frac{\int_0^x\e^{2V(y)/\kappa}y^{-2\epsilon_1/\kappa}(1-y)^{-2\epsilon_2/\kappa}\rd y}{\int_0^1\e^{2V(y)/\kappa}y^{-2\epsilon_1/\kappa}(1-y)^{-2\epsilon_2/\kappa}\rd y}=\int_0^x\e^{\frac{2}{\kappa}\left(V_\eff(y)-\left\langle V_\eff\right\rangle_\kappa\right)}\rd y\ ,
\]
is a stationary solution of the adjoint Kimura equation with mutations
\[
 \partial_t\varphi =\frac{\kappa}{2}x(1-x)\left[\partial_x^2\varphi -V_\eff'(x)\partial_x\varphi\right], 
\]
with $\varphi^{(\epsilon_1,\epsilon_2)}(0)=0$ and $\varphi^{(\epsilon_1,\epsilon_2)}(1)=1$. It is possible to show that $ \lim_{\epsilon_1,\epsilon_2\downarrow 0}\varphi^{(\epsilon_1,\epsilon_2)}(x)=\varphi(x)$,
the fixation probability of the  model without mutation, as defined in Eq.~\ref{eq:fix_prob_continous}. Furthermore, the analysis of $\varphi^{(\epsilon_1,\epsilon_2)}$ will follow closely what as described before for $\varphi$, i.e., assuming small $\kappa$, $\varphi^{(\epsilon_1,\epsilon_2)}$ will be approximately constant whenever $V(x)<\left\langle V_\eff\right\rangle_\kappa$ and it will sharply increase otherwise. 

On the other hand, the stationary distribution of the mutation-Kimura equation is given by
\begin{equation*}
p_\infty^{(\epsilon_1,\epsilon_2)}(x)=C\e^{-2V(x)/\kappa} x^{2\epsilon_1/\kappa-1}(1-x)^{2\epsilon_2/\kappa-1}=C\frac{\e^{-2V_\eff(x)/\kappa}}{x(1-x)},
\end{equation*}	
where 
\[
 C=\left[\int_0^1\frac{\e^{-2V_\eff(x)/\kappa}}{x(1-x)}\rd x\right]^{-1}
\]
is a normalising constant. Note that $p^{(\epsilon_1,\epsilon_2)}_\infty\in L^1([0,1])$ for all $\epsilon_{1,2}>0$, but $\lim_{\epsilon_{1,2}\downarrow 0} p^{(\epsilon_1,\epsilon_2)}_\infty\not\in L^1([0,1])$. Indeed, it can be show that, in an appropriate sense, it will converge to any convex  combination of Dirac masses supported at the endpoints of the interval [0,1] --- see \cite{ChalubSouza09b,ChalubSouza14a} for a discussion on how to make sense of this measure as a stationary solution of the KE.

For the sake of simplicity of presentation, we will resort to a family of quadratic coexistence games, $V_{x_*}(x)=-xx_*+x^2/2$. In these cases, the potential has a unique interior minimum, but only if $x_*=1/2$ this minimum is an $\ess_\kappa$ for all values of $\kappa$. Therefore, in the absence of mutations, only in this case we expect that the system will stay close to the inner equilibrium for a long time. 

\begin{figure}
\includegraphics[width=0.32\textwidth]{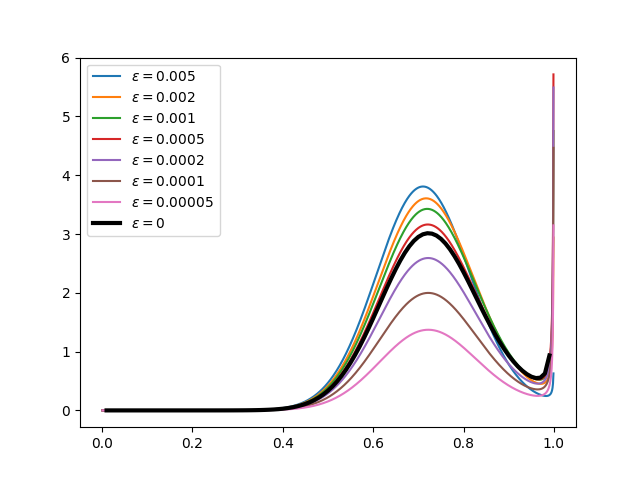}
\includegraphics[width=0.32\textwidth]{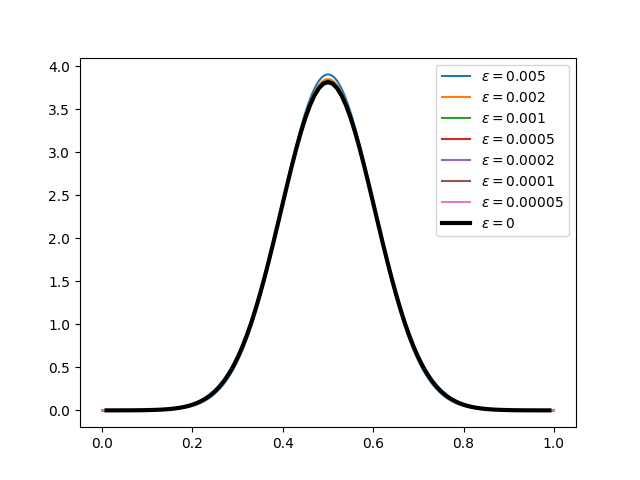}
\includegraphics[width=0.32\textwidth]{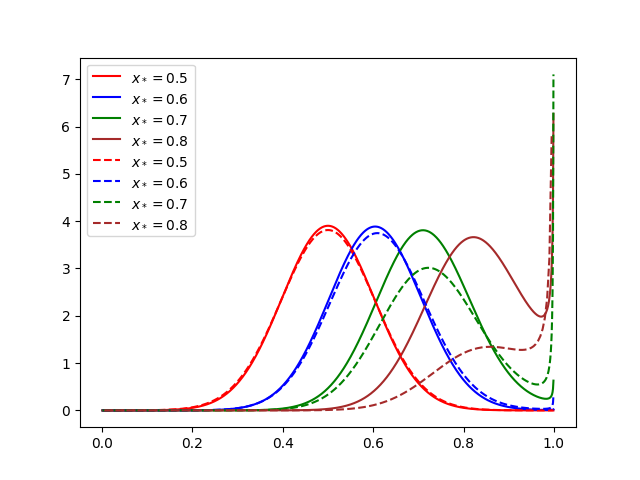}
\caption{ Left and center: The quasi-stationary distribution $p^{(\epsilon,\epsilon)}_\infty$ for several values of $\epsilon$ (different colors) and $\epsilon=0$ (black), for $x_*=0.7$ (left) and $x_*=1/2$ (center).
% for $V_1$ (left) and $V_2$ (center). 
Note that in the second case, where the minimum of the fitness potential is an $\ess_\kappa$, $p^{(\epsilon,\epsilon)}_\infty$ is approximately $\epsilon$-independent, which does not occur in the first case. Right: the value of the quasi-stationary distribution for a symmetric mutation rate $\epsilon=0.005$ (continuous line) and no-mutation $\epsilon=0$ (dashed line), for different values of $x_*$ in the family of potentials $V_{x_*}(x)=-x_*x+x^2/2$, indicated by different colors. Note that $x_*$ is the minimum of the potential, and therefore the maximum of $p^{(\epsilon,\epsilon)}_\infty$, but only $x_*=0.5$ is an $\ess_\kappa$ for any value of $\kappa$. These results suggest that when $x^*$ is an ESS$_\kappa$, the QS distribution is quite robust against sufficiently small mutation rates --- meaning that the stationary distribution with small mutation is very similar to the QS distribution in the interior. On the other hand, when $x^*$ is not an ESS$_\kappa$ small mutations produces stationary distributions that are significant different in the interior. In particular, the mass in the interior is much smaller than the mass in the endpoints.}
\label{fig:mutdependence}
\end{figure}

\begin{figure}
\includegraphics[height=0.28\textwidth]{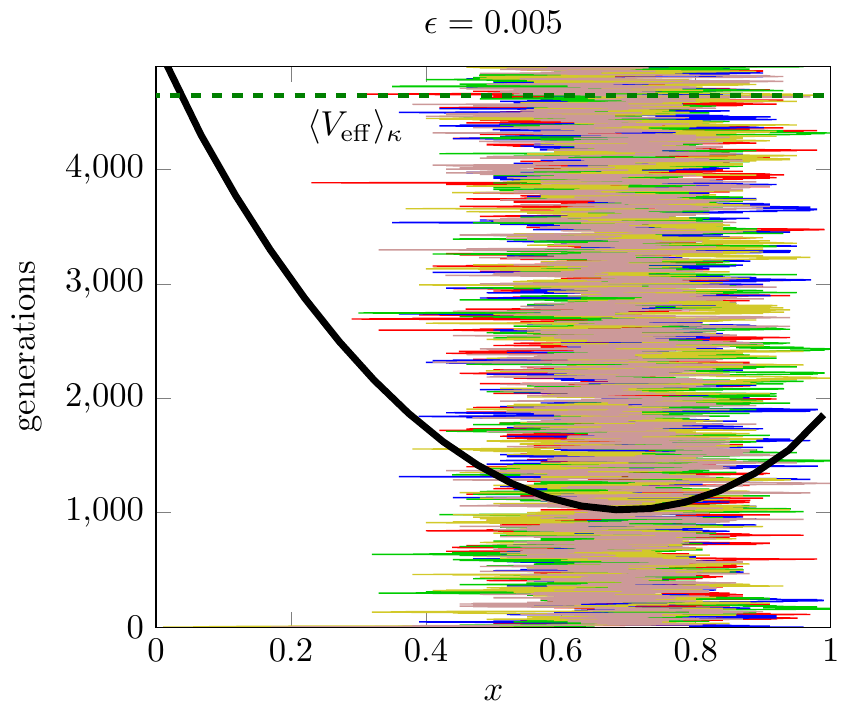}%
\includegraphics[height=0.28\textwidth]{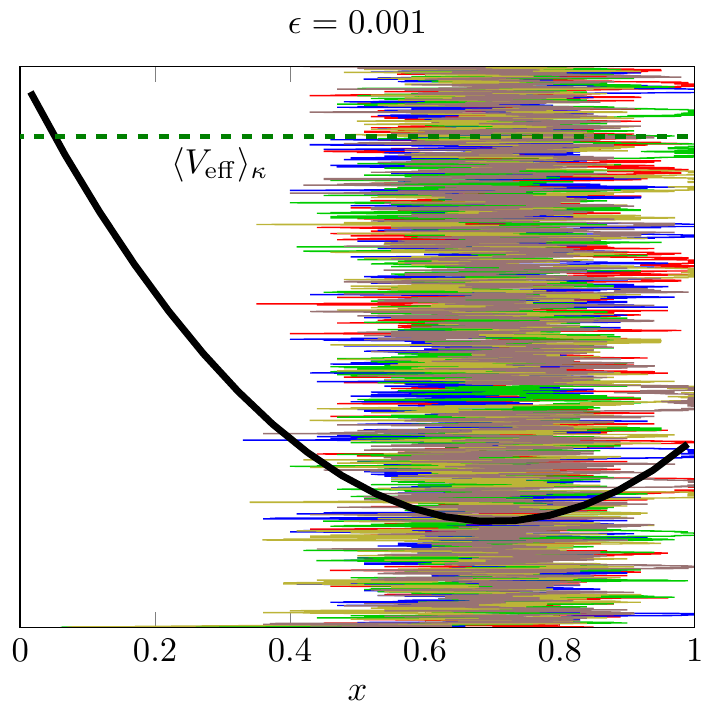}%
\includegraphics[height=0.28\textwidth]{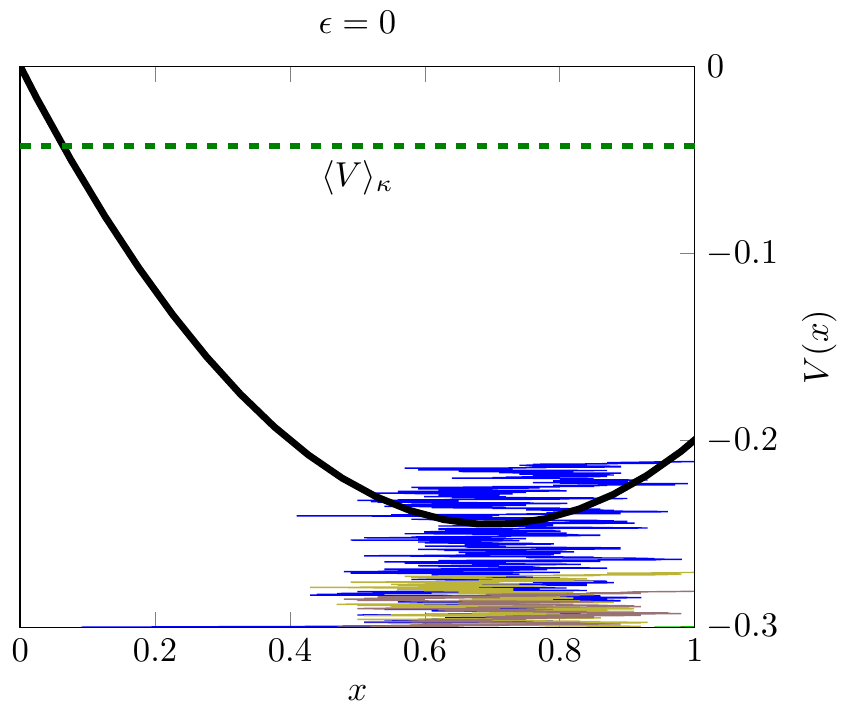}%
\\
\includegraphics[height=0.28\textwidth]{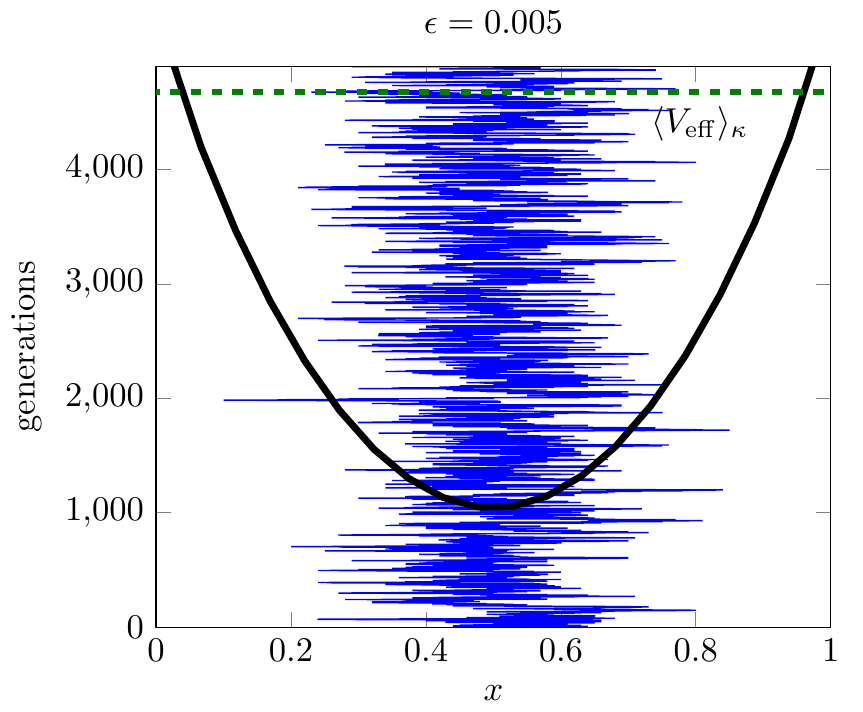}%
\includegraphics[height=0.28\textwidth]{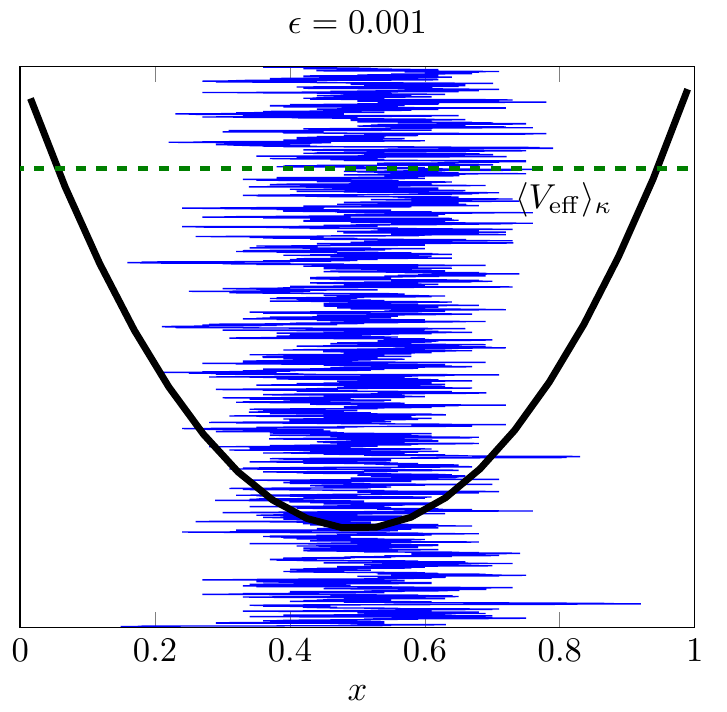}%
\includegraphics[height=0.28\textwidth]{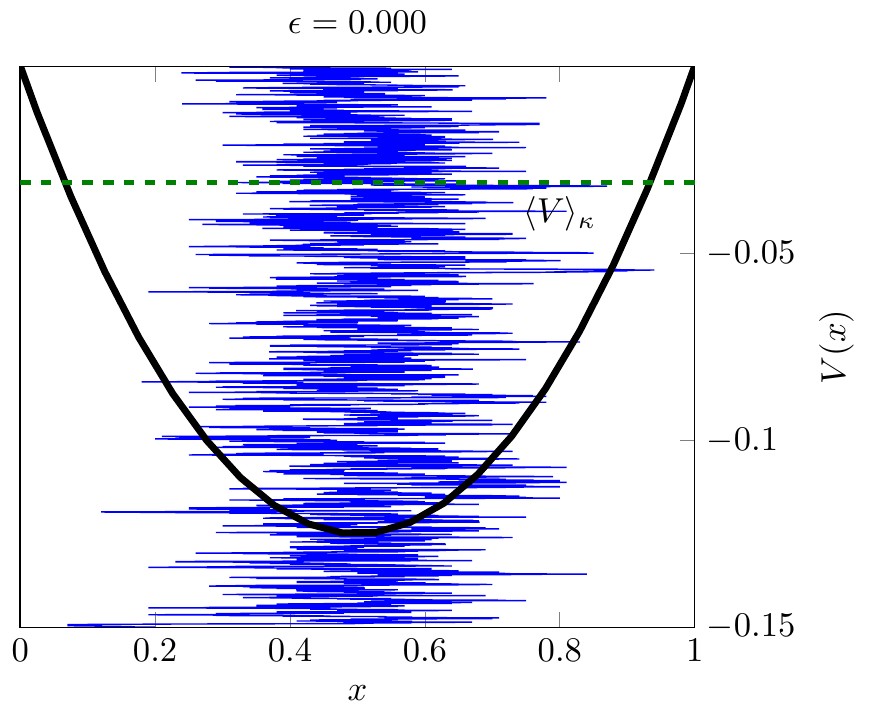}%

\caption{First row: Five simulations (different colors) for the WF process with mutation ($\epsilon=0.005$, $0.001$, left and center columns) and without mutation ($\epsilon=0$, right column). Note that while in the last case, typically the system converges to fixation, as predicted, in the two previous cases the system persists much longer around the minimum of the potential $V(x)=-0.7x+x^2/2-\epsilon(\log x+\log(1-x))$, with weak selection, $\kappa=0.02$ and $N=100$. In the mutation case, the stationary distribution is centered at the only minimum of the potential, $x_*\approx 0.0691$, $0.698$ and $0.7$ for $\epsilon=0.005$, $0.001$ and $0$, respectively. This potential is exactly equivalent to the coexistence game studied in~\cite[Fig. 1]{Vasconcelos_et_al_PRL2017}. The last row shows one simulation, with different mutation rates for the symmetric potential $V(x)=-x(1-x)/2$, associated to a coexistence game with equilibrium at $x_{\min}=1/2$. Note that in this case $x_{\min}$ is an $\ess_\kappa$. The only significant change in the WF behaviour is in the no-mutation case.}
\label{fig:mutation}
\end{figure}

When mutation is considered, there is an effective repulsive force close to $x=0$, $x=1$ such that whenever the population is close to  a monomorphic state the system might go back to the interior minimum. In our mechanical analogy, we say that there is not enough energy to reach fixation. 
Nevertheless, while the system will stay close to minimum for $V_{1/2}$, this will not be the case for $V_{x_*}$, $|x_*-1/2|>\mathcal{O}(\kappa)$. 
% Nevertheless, while the system will stay close to minimum for $V_2$, this will not be the case for $V_1$. 
Indeed, computing the stationary distribution of the WF process shows that when $x^*$ is an ESS$_\kappa$ we find that, as long as the mutation is small, but not too small, the stationary distribution is well approximated by the quasi-stationary distribution of the non-mutation case. Moreover, the stationary distribution has a significant mass at the endpoints --- see Fig.~\ref{fig:mutdependence}. See~\cite{Vasconcelos_et_al_PRL2017} to a detailed study of this example, and see Fig.~\ref{fig:mutation} for numerical simulations of small mutation and no-mutation in the coexistence case.

Furthermore, the behaviour of the stationary distribution in the no-mutation limit seems also to depend on whether the minimum of $V$ is an $\ess_\kappa$. Namely, it seems that the introduction of a small mutation rate has little effect in the stationary distribution only if no interior equilibrium is an $\ess_\kappa$; c.f. Fig~\ref{fig:mutdependence}. In particular, these results are compatible with the corresponding approximation results in \cite{wu2012small}.

\subsection{Discussion}

Starting from a re-interpretation of the fixation probability formula yielded by the standard diffusion approximation as work integrals, we derive a very effective procedure to provide a 
% qualitative 
description of the fixation of probability under the WF dynamics. In this vein, this confirms that the important object seems to be the fitness potential and not the selection gradient --- as already suggested by some of the results in \cite{ChalubSouza16a}.  While this procedure is geared towards populations with two types (2T), we provide  numerical examples for a population consisting of three types (3T), showing that the interpretation of the system moving along lines that require minimum potential is consistent with the multi-type evolution until the first extinction, that will eventually happen with probability $1$.

Essentially, this follows from the fact that any finite population will eventually reach a monomorphic state. Extinctions will be typically sequential, and thus it is sufficient to build a qualitative theory only until the first extinction, as the subsequent evolution will be described by the same theory with one type less. Evidently,   the topology of the $n-1$-dimensional simplex, for $n\ge 3$, is substantially more complicated than when $n=2$, and therefore our picture is far from being complete. However, Figs.~\ref{fig:3T_first_extinction},~\ref{fig:3T_esskappaA} and~\ref{fig:3T_esskappaB} show that the same heuristics for the $\ess_\kappa$ seem to apply to this more general case.

Furthermore, using numerical simulations, we also show that key features of our work seems to be valid in more general situation, e.g., when mutations are introduced in a 2T problem. In particular, we showed that an $\ess_\kappa$ is robust against small perturbation, however, the existence of stable or quasi-stable states in the mutation case is a far more general phenomena than in the no-mutation case. This comes as no surprise, as  many refinements of Nash equilibrium (NE) and evolutionary stable strategies were introduced to cope with the fact that frequently NE are not robust against changes of the game (set of pure strategies, pay-off, or error in the execution of a given strategy). In some of these refinements (e.g.:  the \emph{trembling hand perfection}), to each interior NE of the game, there is a NE of the perturbed game; the same is not true for NE on the boundaries. In other refinements (e.g.; the \emph{essential} NE) existence of NE of the perturbed game is not guaranteed even close to interior NE of the original game. See~\cite{Weibull} for a comprehensive discussion in Nash-equilibrium refinements.

The introduction of mutations in the model, even small ones, may have a profound impact in the overall dynamics. This can be seen, for example, in the use of the Perron-Frobenius theorem to quantify stationary and quasi-stationary distributions of the evolutionary process. If there are no-mutations, the Markov chain in reducible, the leading eigenvalue is doubly-degenerated and the quasi-stationary distribution is given by the eigenvector associated to the sub-leading eigenvalue. The fixation probability is a leading eigenvector of the backward evolution~\cite{ChalubSouza17a}. On the other hand, if mutations are introduced, even small ones, the associated Markov chain is irreducible, the leading eigenvalue is simple, the fixation probability vector is not defined and the stationary distribution is the unique (up to normalization) leading eigenvector.

Note that we loosely defined the idea of ``work necessary to transport a given type from the current state towards fixation''. From the mechanistic analogy that inspired this work, and in particular from the  well-known ``work-energy theorem'', we may define an internal energy associated to each type: we say that the internal energy of type, say, \A is zero if the population is monomorphic at type \A and at state $x$ is given by the minimum amount of work necessary to go from $x$ to fixation, i.e., $\mathcal{E}(x)=\int_x^1\e^{\frac{2}{\kappa}(V(y)-\Vmean)}\rd y$. Therefore, the ratio of the fixation probability of types \A and \B is the inverse of the ratio of the internal energies of types \A and \B. We conclude that the fixation probability of a given type is inversely proportional to the internal energy, with a constant of proportionality that is type-independent (but may depend on $x$). A simple calculation shows that $\varphi^{(\Z)}(z)=\rho(z)\mathcal{E}^{(\Z)}(x)$, of $z$ denotes the presence of the focal type $\Z=\A,\B$. This can also be seen from the fact that $\rho(x)$ is invariant under the change $\A\leftrightarrow\B$. The next step in such a mechanistic formulation of the WF process would be the generalization of this reasoning to multi-type evolution. However, for $n\ge 3$ the infinitude of paths from $x$ to any given point on the boundary of the simplex presents a challenge yet to be overcome.

Finally, we hope that the method presented here can contribute to a more detailed understanding of evolutionary models for multi-player games in finite populations~\cite{Kurokawa:Ihara:2009,Gokhale:Traulsen:2010,Kurokawa:Ihara:2013}. 

\footnotesize 

%\paragraph{Competing interests}
%
%The authors declare no conflict of interest. 
%
%\paragraph{Authors' contributions}
%
%Both authors  contributed equally and approved the current version of the manuscript.

\paragraph{Acknowledgements}

FACCC was partially supported by FCT/Portugal Strategic Project UID/MAT/00297/2013 (Centro de Matemática e Aplicações, Universidade Nova de Lisboa) and by a ``Investigador FCT'' grant. MOS was partially supported by CNPq under grants  \# 486395/2013-8 and  \# 309079/2015-2. We also thank two anonymous reviewers, whose comments helped us to improve the paper. In particular, one of the reviewers pointed out to us the potential connection to mutation models.

\bibliographystyle{vancouver}
\bibliography{fitpot}

\begin{thebibliography}{10}

\bibitem{Imhof:Nowak:2006}
Imhof LA, Nowak MA.
\newblock {Evolutionary game dynamics in a Wright-Fisher process}.
\newblock J Math Biol. 2006;52(5):667--681.

\bibitem{Moran}
Moran PAP.
\newblock The Statistical Process of Evolutionary Theory.
\newblock Oxford: Clarendon Press; 1962.

\bibitem{Fisher1}
Fisher RA.
\newblock On the dominance ratio.
\newblock Proc Royal Soc Edinburgh. 1922;42:321--341.

\bibitem{Wright2}
Wright S.
\newblock The distribution of gene frequencies in populations.
\newblock Proc Nat Acad Sci USA. 1937;23:307--320.
\newblock Available from: \url{http://www.jstor.org/stable/87433}.

\bibitem{ChalubSouza17a}
{Chalub} FACC, {Souza} MO.
\newblock {On the stochastic evolution of finite populations}.
\newblock J Math Biol. 2017 Feb;75(6-7):1735--1774.

\bibitem{ChalubSouza09a}
Chalub FACC, Souza MO.
\newblock From discrete to continuous evolution models: a unifying approach to
  drift-diffusion and replicator dynamics.
\newblock Theor Pop Biol. 2009;76(4):268--277.
\newblock Also available as a Arxiv preprint: 0811.0203.

\bibitem{ChalubSouza14a}
Chalub FACC, Souza MO.
\newblock {The frequency-dependent Wright-Fisher model: diffusive and
  non-diffusive approximations}.
\newblock J Math Biol. 2014;68(5):1089--1133.

\bibitem{Goldstein}
Goldstein H.
\newblock Classical Mechanics.
\newblock 2nd ed. Reading: Addison-Wesley; 1980.

\bibitem{Charlesworth}
Charlesworth B, Charlesworth D.
\newblock Elements of Evolutionary Genetics.
\newblock Greenhood Village, Colorado: Roberts and Company Publishers; 2010.

\bibitem{EthierKurtz}
Ethier SN, Kurtz TG.
\newblock Markov Processes: Characterization and Convergence.
\newblock Wiley Series in Probability and Mathematical Statistics: Probability
  and Mathematical Statistics. New York: John Wiley \& Sons Inc.; 1986.
\newblock Characterization and convergence.

\bibitem{ChalubSouza09b}
Chalub FACC, Souza MO.
\newblock A non-standard evolution problem arising in population genetics.
\newblock Communications in Mathematical Sciences. 2009;7(2):489--502.
\newblock Also available as an Arxiv preprint.

\bibitem{ChalubSouza16a}
Chalub FACC, Souza MO.
\newblock Fixation in large populations: a continuous view of a discrete
  problem.
\newblock J Math Biol. 2016;72(1-2):283--330.
\newblock Available from: \url{http://dx.doi.org/10.1007/s00285-015-0889-9}.

\bibitem{Carvalho_2016}
de~Carvalho M.
\newblock Mean, What do You Mean?
\newblock The Am Stat. 2016;70(3):270--274.
\newblock Available from:
  \url{http://dx.doi.org/10.1080/00031305.2016.1148632}.

\bibitem{Nowak:06}
Nowak MA.
\newblock Evolutionary Dynamics: {E}xploring the Equations of Life.
\newblock Cambridge, MA: The Belknap Press of Harvard University Press; 2006.

\bibitem{smith1973logic}
Smith JM, Price GR.
\newblock The logic of animal conflict.
\newblock Nature. 1973;246(5427):15.

\bibitem{may1994superinfection}
May RM, Nowak MA.
\newblock Superinfection, metapopulation dynamics, and the evolution of
  diversity.
\newblock Journal of Theoretical Biology. 1994;170(1):95--114.

\bibitem{may1995coinfection}
May RM, Nowak MA.
\newblock Coinfection and the evolution of parasite virulence.
\newblock Proc R Soc Lond B. 1995;261(1361):209--215.

\bibitem{kerr2002local}
Kerr B, Riley MA, Feldman MW, Bohannan BJ.
\newblock Local dispersal promotes biodiversity in a real-life game of
  rock--paper--scissors.
\newblock Nature. 2002;418(6894):171.

\bibitem{nowak2002computational}
Nowak MA, Komarova NL, Niyogi P.
\newblock Computational and evolutionary aspects of language.
\newblock Nature. 2002;417(6889):611.

\bibitem{trivers1971evolution}
Trivers RL.
\newblock The evolution of reciprocal altruism.
\newblock The Quarterly review of biology. 1971;46(1):35--57.

\bibitem{axelrod1981evolution}
Axelrod R, Hamilton WD.
\newblock The evolution of cooperation.
\newblock science. 1981;211(4489):1390--1396.

\bibitem{boyd2005origin}
Boyd R, Richerson PJ.
\newblock The origin and evolution of cultures.
\newblock Oxford University Press; 2005.

\bibitem{nowak2005evolution}
Nowak MA, Sigmund K.
\newblock Evolution of indirect reciprocity.
\newblock Nature. 2005;437(7063):1291.

\bibitem{fudenberg2006imitation}
Fudenberg D, Imhof LA.
\newblock Imitation processes with small mutations.
\newblock Journal of Economic Theory. 2006;131(1):251--262.

\bibitem{wu2012small}
Wu B, Gokhale CS, Wang L, Traulsen A.
\newblock How small are small mutation rates?
\newblock Journal of mathematical biology. 2012;64(5):803--827.

\bibitem{Karlin_Taylor_first}
Karlin S, Taylor HM.
\newblock A first course in stochastic processes.
\newblock {S}econd ed. Academic Press [A subsidiary of Harcourt Brace
  Jovanovich, Publishers], New York-London; 1975.

\bibitem{Ewens_Book}
Ewens WJ.
\newblock Mathematical population genetics. {I}. vol.~27 of Interdisciplinary
  Applied Mathematics.
\newblock 2nd ed. Springer-Verlag, New York; 2004.
\newblock Theoretical introduction.
\newblock Available from: \url{https://doi.org/10.1007/978-0-387-21822-9}.

\bibitem{Hinch1991}
Hinch EJ.
\newblock Perturbation Methods.
\newblock U.K.: Cambridge University Press; 1991.

\bibitem{ChalubSouza_ArXiv2018}
{Chalub} FACC, {Souza} MO.
\newblock {From fixation probabilities to d-player games: an inverse problem in
  evolutionary dynamics}.
\newblock ArXiv e-prints. 2018 Jan;1801.02550.

\bibitem{HofbauerSigmund}
Hofbauer J, Sigmund K.
\newblock Evolutionary Games and Population Dynamics.
\newblock Cambridge, UK: Cambridge University Press; 1998.

\bibitem{Vasconcelos_et_al_PRL2017}
Vasconcelos VV, Santos FP, Santos FC, Pacheco JM.
\newblock Stochastic Dynamics through Hierarchically Embedded Markov Chains.
\newblock Phys Rev Lett. 2017 Feb;118:058301.
\newblock Available from:
  \url{https://link.aps.org/doi/10.1103/PhysRevLett.118.058301}.

\bibitem{Weibull}
Weibull JW.
\newblock Evolutionary game theory.
\newblock Cambridge, MA: MIT Press; 1995.
\newblock With a foreword by Ken Binmore.

\bibitem{Kurokawa:Ihara:2009}
Kurokawa S, Ihara Y.
\newblock Emergence of cooperation in public goods games.
\newblock P Roy Soc B--Biol Sci. 2009;276(1660):1379--1384.

\bibitem{Gokhale:Traulsen:2010}
Gokhale CS, Traulsen A.
\newblock Evolutionary games in the multiverse.
\newblock P Natl Acad Sci USA. 2010;107(12):5500--5504.

\bibitem{Kurokawa:Ihara:2013}
Kurokawa S, Ihara Y.
\newblock Evolution of social behavior in finite populations: a payoff
  transformation in general n-player games and its implications.
\newblock Theor Popul Biol. 2013 Mar;84:1--8.

\end{thebibliography}

\end{document}